\documentclass[usenatbib,usegraphicx]{mn2e}
\usepackage{times}
\usepackage{graphicx, epsfig}
\usepackage[fleqn]{amsmath}
\usepackage{amsfonts}
\usepackage{amssymb}
\usepackage[totalwidth=17.8cm, totalheight=24.0cm]{geometry}

\title[Supermassive black holes with LGS AO]{Determination of masses of the central black holes in NGC524 and NGC2549 using Laser Guide Star Adaptive Optics}

\author[Davor Krajnovi\'c et al.]  {Davor
  Krajnovi\'c,$^1$\thanks{E-mail: dxk@astro.ox.ac.uk} Richard M. McDermid,$^2$
  Michele Cappellari,$^1$
 Roger L.\ Davies$^1$ \\
 $^1$Sub-department of Astrophysics, University of Oxford, Denys Wilkinson Building, Keble Road, OX1 3RH, UK\\
$^2$Gemini Observatory, Northern Operations Center, 670 N. A'ohoku Place,Hilo, Hawaii, 96720, USA\\
}

\pagerange{\pageref{firstpage}--\pageref{lastpage}}
\pubyear{2007}

\def\aj{AJ}             
\def\araa{ARA\&A}       
\def\apj{ApJ}           
\def\apjl{ApJ}          
\def\apjs{ApJS}         
\def\aap{A\&A}          
\def\mnras{MNRAS}       
\def\pasp{PASP}         
\def\nat{Nature}        

\begin{document}
\label{firstpage}
\maketitle

\begin{abstract}
We present observations of early-type galaxies NGC524 and NGC2549 with laser guide star adaptive optics (LGS AO) obtained at GEMINI North telescope using the NIFS integral field unit (IFU) in the {\it K} band. The purpose of these observations is to determine high spatial resolution stellar kinematics within the nuclei of these galaxies and, in combination with previously obtained large scale observations with the SAURON IFU, to determine the masses (M$_{\bullet}$) of the supermassive black holes (SMBH). The targeted galaxies were chosen to have central light profiles showing a core (NGC524) and a cusp (NGC2549), to probe the feasibility of using the galaxy centre as the natural guide source required for LGS AO. We employ an innovative technique where the focus compensation due to the changing distance to the sodium layer is made `open loop', allowing the extended galaxy nucleus to be used only for tip-tilt correction. The data have spatial resolution of $0\farcs23$ and $0\farcs17$ FWHM, where at least $\sim40\%$ of flux comes within 0\farcs2, showing that high quality LGS AO observations of these objects are possible. The achieved signal-to-noise ratio (S/N$\sim50$) is sufficiently high to reliably determine the shape of the line-of-sight velocity distribution. We construct axisymmetric three-integral dynamical models which are constrained with both the NIFS and SAURON data. The best fitting models yield  M$_{\bullet}$=($8.3^{+2.7}_{-1.3}  )\times 10^8$ M$_{\sun}$ and $(M/L)_I=5.8\pm 0.4$ for NGC524 and M$_{\bullet}$=($1.4^{+0.2}_{-1.3} )\times 10^7$ M$_{\sun}$ and $(M/L)_R=4.7\pm 0.2$ for NGC2549 (all errors are at the $3\sigma$ level). We demonstrate that the wide-field SAURON data play a crucial role in the M/L determination increasing the accuracy of  M/L by a factor of at least 5, and constraining the upper limits on black hole masses. The NIFS data are crucial in constraining the lower limits of  M$_{\bullet}$ and in combination with the large scale data reducing the uncertainty by a factor of 2 or more.  We find that the orbital structure of NGC524 shows significant tangential anisotropy, while at larger radii both galaxies are consistent with having almost perfectly oblate velocity ellipsoids. Tangential anisotropy in NGC524 coincides with the size of SMBH sphere of influence and the core region in the light profile. This agrees with predictions from numerical simulations where core profiles are the result of SMBH binaries evacuating the centre nuclear regions following a galaxy merger. However, being a disk dominated fast rotating galaxy, NGC524 has probably undergone through a more complex evolution. We test the accuracy to which M$_{\bullet}$ can be measured using seeings obtained from typical LGS AO observations, and conclude that for a typical conditions and M$_{\bullet}$ the expected uncertainty is of the order of 50\%.

\end{abstract}

\begin{keywords} galaxies: individual (NGC524, NGC2549)- galaxies: elliptical and lenticular - galaxies:
kinematics and dynamics - integral field spectroscopy
\end{keywords}

%
%

\section{Introduction}
\label{s:intro}

The discovery that the masses of central dark objects correlate with the global characteristics of their host galaxies, such as the total luminosity or mass\citep{1995ARA&A..33..581K,1998AJ....115.2285M}, led to a realisation that these central objects are crucial for our interpretation of galaxy formation and evolution \citep[e.g.][]{1998A&A...331L...1S, 2005Natur.433..604D, 2006MNRAS.365...11C,2006MNRAS.370..645B,2006ApJS..163....1H}.  The mass of these dark objects, and their link with active galactic nuclei suggest they are supermassive black holes (SMBHs). The most secure evidence for the existence of a nuclear SMBH comes from the centre of the Milky Way, where the large mass is determined by the three-dimensional orbits of stars rotating around it \citep[e.g.][]{2002Natur.419..694S,2003ApJ...586L.127G,2008ApJ...689.1044G,2009ApJ...692.1075G}. Its small physical size is confirmed by the variation of spatially unresolved activity which is linked to an accretion disk around the object \citep[e.g.][]{2001AJ....121.2610D,2005Natur.438...62S,2009arXiv0901.3723F}.

Such detailed information from close to the black hole is, however, not available for more distant non-active galaxies which exhibit the scaling relations between the  M$_{\bullet}$ and the host galaxy properties. The masses of these SMBHs can be determined using dynamical tracers such as a disk of gas clouds in Keplerian rotation around the SMBH or the line-of-sight velocity distribution (LOSVD) of stars coupled with dynamical models, which account for the complexity of possible stellar distribution functions.  Both methods require observations of high spatial resolution since the models have to be constrained by resolved kinematics which probe the sphere of gravitational influence of the SMBH (R$_{sph}$), defined as the distance from the black hole at which the potential of the galaxy and SMBH are approximately equal. This spatial scale is usually defined as R$_{sph}= GM_{\bullet}/\sigma^2$, where $\sigma$ is the velocity dispersion of the galaxy (usually measured over a large aperture), and is typically significantly less than an arcsecond in apparent size, even for nearby galaxies.

While R$_{sph}$ is not a hard limit, and the influence of the SMBH will be felt to a lessening degree at larger radii, it does provide a useful estimate of the scale on which the observable effects of an SMBH on the galaxy's dynamics will be most significant. Until recently, only long-slit observations from Hubble Space Telescope (HST) were able to achieve this resolution. These observations greatly improved the accuracy of the M$_{\bullet}$ determinations and led to a discovery of even tighter relations between M$_{\bullet}$ and $\sigma_e$ \citep{2000ApJ...539L...9F,2000ApJ...539L..13G} followed by similar correlations between M$_{\bullet}$ and host galaxy properties \citep{2001ApJ...563L..11G,2002ApJ...578...90F,2003ApJ...589L..21M,2004ApJ...604L..89H,2007ApJ...665..120A, 2009ApJ...691L.142K}.

Even the HST observations, however, have their limits. The HST is a 2.4 m telescope, has only a long-slit spectrograph and it is optimised to work at optical wavelengths. These specifications prohibit the use of the telescope for measuring spatially resolved kinematics in small, faint galaxies, in low surface brightness giant galaxies with core-like nuclear profiles  and in galaxies with dust-obscured nuclei \citep{2003ASPC..291..196F,2005SSRv..116..523F}. Still, a decade of observations with the HST resulted in about 45 secure M$_{\bullet}$ determinations, mostly of early-type galaxies with $\sigma_e=150-350$ km/s and spanning three orders of magnitude in M$_{\bullet}$ \citep{2008PASA...25..167G,2009ApJ...698..198G}.

The advent of adaptive optics (AO) facilities at the 8-10m class of ground based telescopes opens a new door into the study of SMBHs. These observations, whether natural guide star (NGS) or laser guide star (LGS) assisted, are approaching the resolution of the HST at near-infrared wavelengths which are also less affected by dust and, hence, can be used to probe dusty nuclei of spiral galaxies extending the demography of the M$_\bullet-\sigma$ relation  \citep{2006ApJ...643..226H,2006MNRAS.367....2H, 2006ApJ...646..754D, 2007MNRAS.379..909N, 2007ApJ...671.1329N, 2008MNRAS.391.1629N, 2009MNRAS.394..660C}. In addition, several of the instruments behind these AO systems are integral-field units (IFUs), which, given that they map the LOSVD over a two-dimensional area, offer a tighter constraint on the orbital distribution \citep{2002MNRAS.335..517V} needed for a robust recovery of distribution function from the observables \citep{2005CQGra..22S.347C,2005MNRAS.357.1113K,2008MNRAS.385..614V} . 

The link between the SMBHs and the formation of the host, both in terms of mechanisms and the evolution of this link, is still somewhat hidden behind the uncertainties of the M$_\bullet-\sigma$ relations. The shape, extent and scatter of this relation depend not only on the accuracy of the M$_\bullet$ determinations, but also on the usage of a consistent measure of $\sigma$, and the selection biases of the galaxies used in this relation. In this respect, future efforts need to be focused on extending the range of $\sigma$ probed in this relation, ideally for unbiased samples using consistent techniques for measuring M$_\bullet$. This work aims at taking the first step in this direction, refining our ground-based observational techniques to allow larger numbers of objects to be studied in a homogeneous way.

Spectroscopic observations of nuclei can also be used to probe the formation mechanism of SMBHs.  If central SMBHs are ubiquitous in galaxies (or at least present in those with bulges), then when smaller galaxies merge, the two central black holes sink to the bottom of the potential well forming a binary which should also merge over some time. This process can only be completed with the assistance of nearby stars;  the energy and the angular momentum of the binary is transferred to stars approaching the binary on radial orbits, which are then ejected at much higher velocities \citep{1997NewA....2..533Q,2001ApJ...563...34M}. This process leads to dynamical heating of the central regions and a decrease in the nuclear surface brightness profile \citep{2008arXiv0810.1681K}. The `scouring' of the core is used to explain the existence of core-like light profiles \citep{1997AJ....114.1771F}, and nuclear mass deficits in massive ellipticals \citep{2002MNRAS.331L..51M,2008arXiv0810.1681K, 2009ApJ...691L.142K}. In addition to producing a core in the light profile, simulations of this process also predict that the merging of an SMBH binary would leave a clear signature in the orbital distribution: there should be a bias towards tangential orbits in the core region \citep[e.g.][]{1996ApJ...465..527M, 1997NewA....2..533Q,2001ApJ...563...34M}. 

The main prohibitive issue with AO observations is that, even with LGS capabilities, a natural guide star is required for an optimal correction of atmospheric aberrations. There are only a handful of galaxies with a near-resolvable R$_{sph}$ that have a suitable guide star, even at the fainter magnitudes permitted by LGS AO. Moreover, the correction provided by current single reference source, single conjugate AO systems quickly degrades when the off-axis distance of the guide star exceeds the isoplanatic patch size, which is typically much smaller than the effective radius of the galaxy. This means that the use of stars for AO observations of nearby galaxies is not practical for large samples. 

This limitation has prompted innovative use of AO techniques, such as neglecting altogether the low-order corrections provided by the natural tip-tilt  \citep{2008Msngr.131....7D}. In this study, we employ a novel method developed at Gemini Observatory which allows the galaxy centre itself to be used as a reference for tip-tilt correction. With this study, we explore the performance of this technique using two objects with differing central light profiles: NGC524 which has a flattened `core' profile, and NGC2549 which has a steep `cusp' profile. These two representative cases will demonstrate the feasibility of applying this technique to larger samples of objects, which in turn will allow us to tackle the issue of refining and better understanding the connection between SMBHs and their host galaxies.

In Section~\ref{s:obs} we present the observation, in particular the LGS setup and the data used. In Section~\ref{s:red} we describe the data reduction and extraction of the kinematics. In Section~\ref{s:res} we discuss the determination of the PSF and AO correction, while in Section~\ref{s:dyn} we present the dynamical models and the main results of this paper. In Section~\ref{s:discuss} we discuss and in Section~\ref{s:conc} we summarise our results.

%
%

\section{Observations and LGS AO setup}
\label{s:obs}

The results of this work are based on the combination of large-scale integral-field kinematics from the SAURON instrument \citep{2001MNRAS.326...23B} and high spatial resolution integral-field kinematics from NIFS on Gemini North. The SAURON observations are presented in \citet{2004MNRAS.352..721E}. In this work we present the new data obtained with NIFS on Gemini North using laser guide star adaptive optics (LGS AO).

\subsection{LGS AO observations}
\label{ss:lgs}

The Altair LGS system consists of a laser that projects a $\sim 10$ Watt coherent beam at 589 nm into the sky, exciting the sodium atoms at the height of around 90 km in the atmosphere, which then re-emit radiation and glow as a source on which Altair can perform the high-order corrections. A more complete description of the Altair LGS system is given in \citet{2006SPIE.6272E.114B}.

The final improvement of the PSF with the LGS system still depends on the existence of a nearby ($<$ 25\arcsec~for Altair) natural guide star (NGS). This star is needed to correct for the angle-of-arrival variations due to the atmospheric turbulence, usually called the tip-tilt correction, and to measure the absolute focus of the telescope. The first correction arises because the deflection of the beam of the LGS when it travels upwards and downwards through the atmosphere cancel each other, while the beam of the object above the atmosphere is deflected only once, leaving the low-order, tip-tilt, disturbance not detected by the LGS AO system. The second correction is a consequence of the finite and slowly varying distance between the telescope and the sodium layer, which prevents the use of the LGS beacon for determining the focus of the telescope at infinity. Both degradations, tip-tilt and change in focus, can be corrected with an NGS (often called `tip-tilt' star), which can, in general, be further away and a few magnitudes dimmer (up to 3 magnitudes for the Altair LGS system) than for a pure NGS AO observation.

For a given guide-source flux rate, LGS AO typically yields smaller Strehl ratios than NGS AO observations due to the effects of the finite size of the LGS beacon in the atmosphere,  the cone effect and increased sensitivity to the actual seeing, which distorts the upward travelling laser beam. For bright, on-axis sources, the predicted Strehl ratio can be as much as 20\% less with an LGS system compared to NGS \citep{1998MNRAS.295..756L}. Nevertheless, they dramatically extend the useful sky coverage, allowing AO-improved observations of targets for which a bright NGS is unavailable \citep{2004SPIE.5490..331S}.  The Altair systems requires, for a low Strehl observations,  a tip-tilt star of 17.5 (R band) magnitude to be within 25\arcsec from the target, the nucleus of the galaxy in our case. Very few galaxies with R$_{sph} > 0\farcs05$ (typically nearby early-type galaxies) have neighbouring stars on the sky that meet this criterion. Increasing the accessible off-axis distance only degrades the PSF, and the use of fainter stars becomes problematic due to the non-negligible background of the extended galaxy at these magnitudes. This poses a serious limitation on what can be achieved on this topic with existing single-guide source, single conjugate AO systems on large telescopes.

\subsection{The open-loop focus model}
\label{ss:focus}

To avoid the issue of suitable guide stars, it is, in principle, also possible to use the nucleus of the galaxy as a natural guide source for the system, provided it is sufficiently bright and compact. For Altair, the `rule of thumb' is that the nucleus should show a drop of $\geq 1$ magnitude in brightness within the central $\sim 1$\arcsec, which ensures a certain degree of contrast within the field of view of the wave-front sensor. This condition is met for many AGNs or quasars, which have bright, unresolved nuclei \citep[e.g. the case of Cen A,][]{2006ApJ...643..226H}, and many nearby spirals have a suitably bright and distinct central star cluster to allow AO corrections. But for quiescent early-type galaxies, the situation is less obvious. The central light profiles of early-type galaxies show a general change from steeply rising `cusp' profiles at lower masses, towards a flatter central profile that can define a central `core' region \citep{1994AJ....108.1598F,1997AJ....114.1771F}. For the most massive and nearby quiescent galaxies, the core region (inside which the surface brightness changes only little) can be larger than the wave-front sensor field of view, which sees essentially a constant background of light from which no tip-tilt information can be obtained. However, this is only relevant for galaxies at the most massive end of the M$_\bullet-\sigma$ relation. The majority of galaxies have either core radii that are similar to or less than one arcsecond, or exhibit steep, cuspy profiles which should show contrast within the AO wave-front sensor. In order to test what correction of the PSF could be expected for typical early-type galaxies, we chose NGC524 and NGC2549 which both satisfy the basic Altair requirement for nuclear guiding. Specifically, NGC524 has a core-like light profile \citep{1997AJ....114.1771F}, while NGC2549 a cusp-like profile at the HST resolution \citep{2001AJ....121.2431R}.

After initial observations were attempted, it became clear that the chosen nuclei, while suitable for tip-tilt correction, are too faint to be used for constraining the focus, resulting in an unstable focus control loop and subsequently compromised image quality. To avoid this, Gemini staff implemented a procedure by which the focus correction during the science integration is controlled by a geometric function that takes into account the change in the distance to the sodium layer as the telescope position changes. This, so-called `open-loop' focus model, was a pre-existing feature of the LGS-AO system, to reduce the convergence time of the system following large slews during night-time operations. The only time-dependent parameter of this model is the altitude (i.e. absolute height above sea level) of the sodium layer, which is determined immediately before observing the galaxy by `tuning' the LGS AO system using a nearby bright star. When all the control loops of the system (tip-tilt, focus, and LGS) have converged with this reference source, the loops are opened and the science target is acquired. The tip-tilt and LGS control loops are then closed, using the galaxy nucleus and laser beacon respectively as reference sources. The focus loop is left open, being passively controlled by the open-loop model. After approximately one hour of observations of the galaxy nucleus, the bright reference star is re-observed; first with the focus loop left open, and then with the focus loop closed, in order to measure the relative degradation of the PSF due to any departure from the open-loop model during the science integration. Consequently, the LGS AO system is now re-optimised at this stage for further observations.

We note that, whilst this technique was under development at the time our observations were obtained, this mode of operation is now fully supported, as of the 2009A semester.

\begin{figure}
        \includegraphics[width=\columnwidth, bb=30 30 590 310]{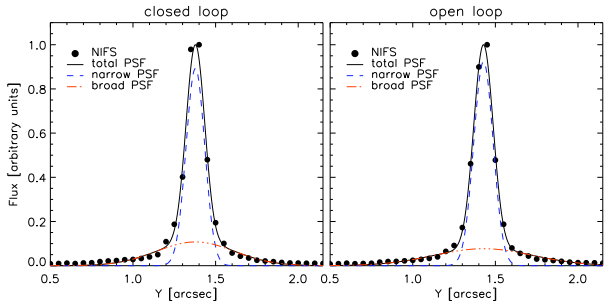}
\caption{\label{f:SFO} The stability of the PSF correction in the open-loop focus model as measured on the horizontal cuts through the reconstructed images of a star used to set up the LGS AO system. On both panels blue and orange lines are the narrow and broad Gaussian components describing the PSF. The red line is the sum of these two components. {\bf Left:} The LGS AO is fully tuned in, with both tip-tilt and focus loops closed and using the star for guiding. The FWHMs are 0\farcs11 and 0\farcs48, for narrow and broad components, respectively. {\bf Right:} One hour later, the LGS AO system is in the same set up as during the observations of the science target: tip-tilt loop is closed on the star, while the focus loop is open and the changes in the focus still follow a model. The degradation is virtually negligible: FWHMs are 0\farcs12 and 0\farcs60 for narrow and broad components, respectively.}
\end{figure}

Figure~\ref{f:SFO} shows the stability of the PSF correction in the open-loop focus model. The panels show horizontal cuts through the reconstructed image of the tuning star observed with LGS AO with an actively optimised focus (left panel) and taken one hour later, where the focus of the telescope has been completely controlled during that time by the open-loop model (right panel). The observed degradation of the PSF is minimal. In both cases, the measured PSFs can be well approximated by a double Gaussian, with full-width half maximum (FWHM) values changing from 0\farcs11 and 0\farcs48 to 0\farcs12 and 0\farcs60 for the narrow and broad components of the PSF, respectively. Approximating the Strehl ratio as the ratio of the total flux in the narrow Gaussian component and the total flux in the whole PSF, we obtain the values of $\sim31$\% for both closed and open loop PSFs. These negligible differences show that it is possible to achieve a very good PSF correction of 0\farcs1-0\farcs2, (FWHM) approaching the diffraction limit in the K band, even without the focus loop closed. On this occasion at least, the Hawaiian skies were stable over approximately one hour, demonstrating that the open-loop focus model used by the Gemini system seems to track accurately the changes in distance to the sodium layer.

The PSFs measured from the tuning star observations, however, cannot be considered representative of the PSF of science observations obtained guiding on the nucleus. The above demonstration reassures us that the change in focus during the galaxy observation was well approximated by the open loop model, and suggests that the delivered PSF of the galaxy centres may also comprise a narrow component comparable to the diffraction limit and a substantial broad component. Estimation of the actual delivered PSF for our galaxies, with their differing central light profiles, is discussed in Section~\ref{s:res}.

\subsection{NIFS observations}
\label{ss:nifsobs}

NGC524 and NGC2549 were observed during the 2007B and 2008A semesters in the K band (central wavelength is $2.2\mu$m) with H+K filter and spectral resolution of R$\sim$5000. The field of view of NIFS is $\sim3\times3\arcsec$. The observations were divided into separate blocks of approximately one hour with 600 seconds science exposures set in the sequence: OSOOSO, where O is an observation of the object (galaxy) and S is an observations of the empty sky field. NGC524 and NGC2549 were observed in total for $21\times600$ and $15\times600$ seconds (3.5 and 2.5 hours) on source, respectively. Some frames had to be, however, discarded in the final stage because of inferior PSF corrections (see Section~\ref{s:res} for more details) and the final data cubes for NGG524 and NGC2549 were constructed out of $10 \times600$ and $11\times600$ on source observations, respectively. In addition to galaxies, we observed telluric standards of A2 V and G0 V types, switching between the stars and the sky fields in the same sequence as for the science targets (see Section~\ref{s:red} for telluric correction).

\subsection{Other data}
\label{ss:other}

In the construction of dynamical models next to kinematic data we also use images of our galaxies. Since we need to constrain the models at the very high spatial resolution in the centre and at large radii in order to collect all available information about the distribution of light, we make use of Hubble Space Telescope (HST) images obtained with Wide Field and Planetary Camera 2 (WFPC2) in  F702W (NGC2549) and F814W (NGC524) bands, and ground based images obtained at the MDM telescope in F814W band (Falc\'on-Barroso et al., in prep).

%
%

\section{Data reduction and stellar kinematics extraction}
\label{s:red}

NIFS data were reduced using the set of reduction routines provided by the GEMINI and incorporated within the IRAF package, following {\it nifsexamples} scripts for the calibration, telluric and science observations\footnote{http://gemini.conicyt.cl/sciops/instruments/nifs/NIFSReduction.html}. Our data reduction departed from the GEMINI guidelines in two instances: at the preparation of the telluric stars and co-addition (merging) of the individual fully reduced (flat fielded, sky subtracted, wavelength calibrated, telluric corrected) science frames. In this section we first describe these two steps and afterwards we outline the method we used for the extraction of kinematics. 

\subsection{Telluric correction}
\label{ss:telluric}

Near-infrared ground-based spectroscopic observations are characterised by the presence of strong atmospheric absorption features. These features exhibit temporal and spatial dependency and have to be removed from the spectra in order to uncover intrinsic spectral characteristics of observed objects. The correction method is based on the fact that by observing a featureless star, both near in time and sky coordinates to the object one studies, one can obtain a reference spectrum of the unwanted absorption lines. While seemingly straightforward, this method is hindered by the lack of objects that are both featureless, reasonably bright and evenly distributed on the sky such that can be found sufficiently near (in air mass) to potential targets. Usually, stars of type A are used since they generally do not have strong metal features, show little reddening and are well approximated by a black body spectrum of 10 000 K. 

\citet{1996AJ....111..537M} outlined how F and G stars can also be used if a high resolution solar spectrum is employed to remove the intrinsic features. In this case a telluric correction spectrum is derived dividing the observed G or F star by a solar template spectrum shifted and broadened to the appropriate relative velocity and resolution. The quality of the correction depends on how similar the telluric star is to the sun. \citet{2003PASP..115..389V} put forward a similar method that is based on A0 V stars, such as Vega. These are a few times more common the G2 V stars and, hence, more likely to be found next to science targets, and can be used to derive a telluric correction spectrum by dividing them by a high-resolution normalised model spectrum of Vega, again shifted and convolved to the resolution of the telluric stars.  
 
We followed these two methods and observed two types of telluric stars: a G0 V Hip3033 during the observations of NGC524 and A2 V Hip44717 and Hip31665 stars for NGC2549. Hip31665 was observed only one night (March 29 2008) covering 4 exposures of NGC2549, the rest begin covered by Hip44717. The telluric stars were reduced as science frames (sky subtracted, flat-fielded, spatially and spectrally  rectified) and one-dimensional spectra were extracted. As in the above studies, we use a high-resolution solar spectrum\footnote{ftp://nsokp.nso.edu/pub/atlas/photatl/} \citep{1991aass.book.....L} and a similarly high-resolution model of Vega\footnote{Obtained from \citet{1991ppag.proc...27K}: http://kurucz.harvard.edu/stars.html} as templates to establish the relative velocity shift and broadening of the telluric stars. We do this using the penalised Pixel Fitting (pPXF) method of \citet[][see Section~\ref{ss:kin} for more details about the method]{2004PASP..116..138C}, which conveniently allows us to select regions of the observed spectrum which are relatively free of telluric absorption, and derive the velocity shift and broadening (described by a truncated Gauss-Hermite polynomial) required to match the star's intrinsic absorption features. Since the high-resolution template spectra are flux calibrated (the normalised solar reference spectrum was renormalized using a black-body temperature of 5800 K), no polynomial term is included in the pPXF, so the residuals of the fits provide both the telluric correction and flux calibration spectrum. We then returned to the GEMINI NIFS pipeline and used the derived telluric correction spectra with task NFTELLURIC to correct each individual data cube for telluric absorption features.

\subsection{Merging of science frames and binning of spectra}
\label{ss:merge}

Before individual science frames can be combined into a final data cube it is necessary to put them on a common wavelength range and sampling, and determine the relative positions (or to re-centre the frames). Frames with bad PSF should also be excluded from the combined data cube, and we discuss this in Section~\ref{s:res}. We re-centred the science frames for each galaxy in the following way: we choose a frame with a reasonable PSF (judged by eye to be one of the best and showing circular and concentric isophotes in the very centre), and compared all other frames to it. The comparison was done on the isophotes of the reconstructed images (created by summing in the spectral direction), such that the isophotes of the re-centred frames overlap, where the goodness of the overlap is judged by eye. Given the variation in the PSF correction, is it not always possible to achieve a perfect match between the isophotes, especially the central ones, where the influence of the seeing is most significant. Centering on the outer isophotes is usually robust, and the typical uncertainty in re-centering was around 0\farcs01. After all frames were re-centred, we merged them into a final data cube. This was done by a custom-made IDL procedure that reads in all individual (re-cantered) frames, determines their spatial extent, defines a new gird of a given pixel size, and interpolates individual frames to this new common grid. The flux values of the final data cube were obtained as the median of the individual cubes. For both galaxies we interpolated to a grid covering $3\arcsec \times 3\arcsec$ with 0\farcs05 square pixels. These pixels over-sample in the cross-slice direction, which is justified by the oversampling provided by our dither pattern, which uses step sizes equivalent to non-integer numbers of slices, and the large number of individual exposures included in the merge. We slightly reduce the sampling along the slices, although the PSF is still adequately sampled.

Extraction of the higher-order moments of LOSVD requires signal-to-noise ratio, $S/N$, of generally high values,  e.g. $S/N \geq 40$  \citep[][]{1993ApJ...407..525V, 1994MNRAS.269..785B,1995AJ....109.1371S}. We use the Voroni binning technique of \citet{2003MNRAS.342..345C} to ensure a minimum $S/N$ whilst optimising the spatial resolution. We first estimate the noise of the unbinned spectra as the standard deviation of the difference between each spectrum and its median smoothed version (over 30 pixels). For both galaxies we require a target S/N of 80, which results in 370 and 377 bins for NGC524 and NGC2549, respectively. In the central 1 arcsecond of the NIFS field-of-view, the data are binned to no larger than $0\farcs1\times0\farcs1$ total size. We checked {\it a posteriori} the achieved S/N in each bin by comparing the median value of a spectrum with the standard deviation of the residuals (difference between the optimal stellar template and the observed spectra, see Section~\ref{ss:kin}). In most cases for both galaxies, the final S/N per bin was in the range from 40-60.

\subsection{Stellar kinematics}
\label{ss:kin}

We use the penalized Pixel Fitting (pPXF) method of \citet{2004PASP..116..138C} to derive the LOSVD in pixel space, parameterised by a Gauss-Hermite polynomial \citep{1993MNRAS.265..213G, 1993ApJ...407..525V}. pPXF finds the best fit to a spectrum by convolving a template spectrum with the corresponding LOSVD given by the mean velocity $V$, velocity dispersion $\sigma$ and Gauss-Hermite moments $h_3$ and higher, which can be used to quantify the asymmetric and symmetric deviations of the LOSVD from a Gaussian.

To avoid template mismatch effects, an `optimal template' spectrum is derived as the linear combination of spectra from a template library which, for a given LOSVD, best represent the galaxy spectrum. Ideally, the template library should contain stars which are representative of the galaxy's stellar population. The K band light of early-type galaxies is dominated by cool and evolved giant stars (red giant and asymptotic giant branch stars). We used the stellar library of \citet[][version 1.0, April 2007]{2008RMxAC..32..177W} publicly available from the GEMINI NIR Resources web site\footnote{http://www.gemini.edu/sciops/instruments/nifs/near-ir-resources}. This library consists of 29 stars covering the spectral types from F7 to M3 and luminosity classes I, II, III and V. We use a subset of these stars that was observed in the (combined) wavelength range from 2.15 to 2.43 $\mu m$. Although the stars were observed with GNIRS, the library contains reduced one dimensional spectra with instrumental dispersion $\sigma_{inst}= 26$ km/s, which compares well with the spectral resolution of our data: $\sigma_{inst}$ = 30 km/s, as measured from the broadening of night-sky emission lines. The publicly available spectra are not corrected to rest wavelength, and must be brought to a common velocity before they can be combined as a velocity template. This was done using pPXF to measure relative shifts between the library stars, using the high-resolution solar spectrum used in Section~\ref{ss:telluric} as a template. The distribution of relative velocities was checked using a star from the library itself as a template, to explore the effects of template mismatch. After de-redshifting the spectra with the solar template, the dispersion in velocities measured using the library template was less then 2 km/s.

In the K band, the stellar kinematics is mostly constrained by the CO band head starting at $\lambda$ = 2.29 $\mu$m and consisting of  $^{12}$CO (2,0), (3,1), (4,2) and (5,3) transitions \citep{1997ApJS..111..445W}. In our spectra the information on the stellar continuum can only be found blueward of the $^{12}$C0 (2,0) transition, and while it is important to keep as much as possible of this region during the extraction, the blue end of the usable spectra is given by the blueward extent of stellar templates. In this largely featureless region possible absorption-lines are Na I, Fe IA, Fe IB, Ca I and Mg I \citep[e.g.][]{2008ApJ...674..194S}. 

Before calling pPXF we rebin all spectra in wavelength to a linear scale x=ln $\lambda$, while preserving the number of spectral pixels, and convolve the template spectra with the (quadratic) difference in the spectral resolution between the NIFS and GNIRS spectra. We describe the LOSVD with the $V$, $\sigma$, $h_3 - h_6$ moments, and add a 4th order additive polynomial to correct for the shape of the continuum, given that  the template stars of \citet{2008RMxAC..32..177W} are already continuum subtracted. Prior to extracting the kinematics, we performed Monte Carlo simulations  to determine the level of penalisation used, and determined that $\lambda=0.4$ is optimal for our data \citep[see definition in ][]{2004PASP..116..138C}. The penalisation reduces the scatter in the derived kinematic parameters when the intrinsic dispersion becomes similar to the instrumental dispersion (and therefore usually critically sampled) by biasing the LOSVD toward a simple Gaussian. In principle, this is more important in the case of NGC2549 than for NGC524, since their $\sigma_e$ are 145 and 235 km/s, respectively \citep{2007MNRAS.379..401E}. In both cases, however, the LOSVDs are expected to be well sampled by the observational set up ($\sigma_e > $ pixel scale in km/s), making the penalisation less important. 

\begin{figure}
        \includegraphics[width=\columnwidth, bb=28 28 595 595]{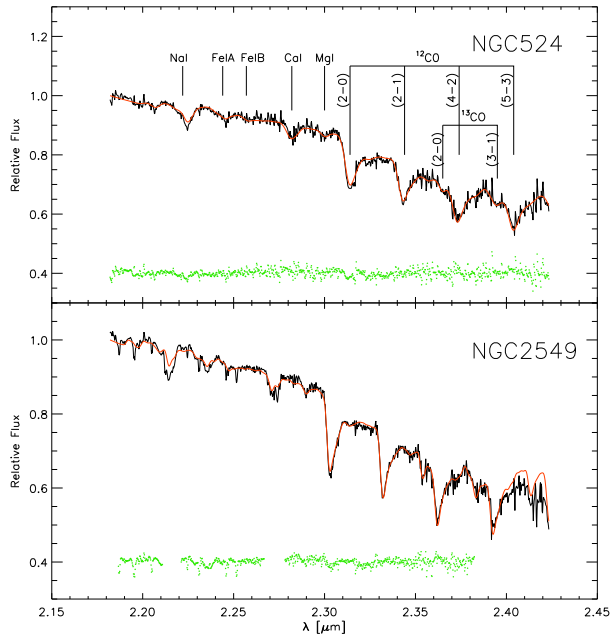}
\caption{\label{f:opttemp} Optimal templates for NGC524 ({\bf top}) and NGC2549 ({\bf bottom}). In both cases, black solid lines show the total galaxy spectrum (sums of spectra of all spatial bins), red lines are the optimal templates, while green dots are the fit residuals, shifted upwards by an arbitrary amount. Regions without residuals were excluded from the fit (only for NGC2549). Vertical lines and associated names on the top panel describe the main absorption lines detectable on spectra of both galaxies. Note that the (5-3) $^{12}$CO feature is fairly well reproduced even if it lies outside the fitting range for NGC2549.}
\end{figure}

Rather than determining an optimal template for every bin, a `global' optimal template was derived for each galaxy, which was then held fixed for each bin in that galaxy (the polynomial is still free to vary - only the relative mixture of spectra from the library is fixed). Each optimal template was determined from a pPXF fit to the sum of all the binned spectra of the galaxy, using the full template library of 23 stars. The resulting optimal templates show certain differences in the stellar populations between the two galaxies. While for NGC524 we considered the full wavelength region common to the stars and the NIFS spectra spanning $2.18-2.42 \micron$, this approach for NGC2549 yielded a considerable template mismatch, mostly visible in the spatially non point-symmetric distribution of $h_3$ values across the field-of-view - often taken as an indicator of template mismatch \citep{1994MNRAS.269..785B}. In addition, (2-0) and (5-3) $^{12}$CO absorption features were not reproduced well, their depths being under- and over-estimated, respectively. In order to minimise this template mismatch, we masked NaI and CaI  absorptions lines and truncated the fitting region before the fourth $^{12}$CO (5-3) transition when fitting the combined spectrum of NGC2549. Figure~\ref{f:opttemp} presents the final optimal templates, the absorption lines used and the residuals to the fits. 

\begin{figure}
        \includegraphics[width=0.9\columnwidth, bb=30 30 390 1020]{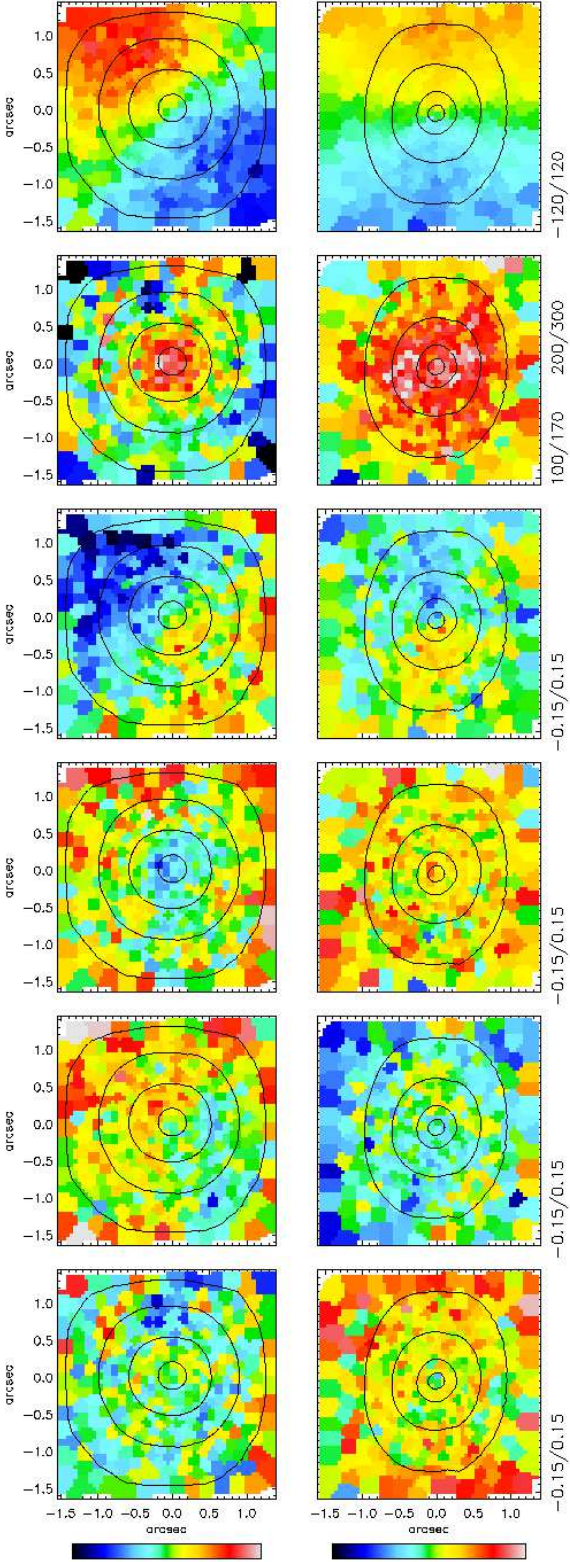}
\caption{\label{f:maps} Kinematics maps of NGC524 {\bf (left)} and NGC2549 {\bf (right)}. From top to bottom maps present: the mean velocity $V$, velocity dispersion $\sigma$, and Gauss-Hermite moments: $h_3$, $h_4$, $h_5$ and $h_6$. The colour-bar at the bottom show the plotted colour range, while the lower and upper limits for each map are given to the right of NGC2549 maps as set of two numbers. In the case of $\sigma$ maps, the upper pair of numbers refers to NGC524 and lower to NGC2549. Note that even $h_5$ and $h_6$, although noisy, retain their main characteristics of being anti-symmetric and symmetric across maps, respectively. North is up and East to the left.}
\end{figure}

\begin{table}
   \caption{Stellar templates used in the construction of optimal templates for NGC524 and NGC2549 from the template library of \citet{2008RMxAC..32..177W}.}
   \label{t:stars}
$$
  \begin{array}{ll|ll}
   \hline
    \noalign{\smallskip}

    \multicolumn{2}{c|}{$NGC524$ } &  \multicolumn{2}{c}{$NGC2549$}\\ 

    star & type & star & type \\
    \noalign{\smallskip} \hline \hline \noalign{\smallskip}
   HD113538\dagger & K8V  &  HD113538  & K8V  \\
   HD32440\dagger   & K6III &  HD2490     & M0III  \\
   HD63425B & K7III & HD4730     & K3III  \\
                    &          &     HD63425B & K7III \\
                    &          &     HD720       & K5III  \\
      \noalign{\smallskip}
    \hline
  \end{array}
$$ 
Notes -- Information on the type was obtained from GEMINI NIR Resources web site. Stars with $\dagger$ are given different type in the Simbad Astronomical Database (http://simbad.u-strasbg.fr/simbad/): HD113538 is K9V and HD324440 is K4III.  
\end{table}

The resulting optimal template for NGC524 combines only three stars from the library, while in the case of NGC2549, a total of five stars were selected by the pPXF fit to form the optimal template. The stars, their types and the weights are presented in Table~\ref{t:stars}. In both cases, a K giant star can reproduce the main spectral features (the CO bandhead) to the first order, but to fit the full spectral range a cool dwarf (HD113538 in the library) is also needed. The dwarf star is particularly important for reproducing spectral features blue-ward from the CO bandhead, especially the prominent Na I ($2.20 \micron$), Ca I ($2.26 \micron$) and Mg I ($2.28 \micron$) absorptions. This star, however, does not reproduce well the CO absorption features. The remaining small discrepancies are visible mostly in the fit to the (2-0) transition, and in the case of NGC524 for the NaI absorption feature, while weak $^{13}$CO features are reproduced well (although they are almost blended with the dominant CO bandhead due to the larger velocity dispersion, and are partially outside the fitting range for NGC2549). 

After the optimal template is constructed, we apply it to the spectra of the individual bins, using the same wavelength regions during the fit as in the construction of the optimal template. While it is possible that there are variation in stellar populations between bins which can not be described by the optimal template, the remaining difference can be well fitted by the additive polynomials alone. We visually inspected all spectra to verify that they are well reproduced by the given optimal templates and we recheck if there are evidence for template mismatch. Finally, we estimate the uncertainties from a set of 100 Monte Carlo simulations, where each spectrum has a an added perturbation consistent with random noise. Since we do not propagate the uncertainties of each spectral pixel in the data cube, the level to which a spectrum is perturbed is given by the robust standard deviation of the difference between the best fit and the original spectrum. During the Monte Carlo calculation the penalisation is turned off to produce realistic uncertainties. In the case of NGC524, typical derived errors are 10 km/s, 13 km/s, 0.03, 0.04, 0.04 and 0.04 for $V$, $\sigma$, $h_3$, $h_4$, $h_5$ and $h_6$, respectively. Similarly, for NGC2549 typical errors are: 5 km/s, 7 km/s, 0.03, 0.03, 0.04 and 0.04 for $V$, $\sigma$, $h_3$, $h_4$, $h_5$ and $h_6$, respectively. 

\begin{figure}
        \includegraphics[width=\columnwidth, bb=40 40 590 590]{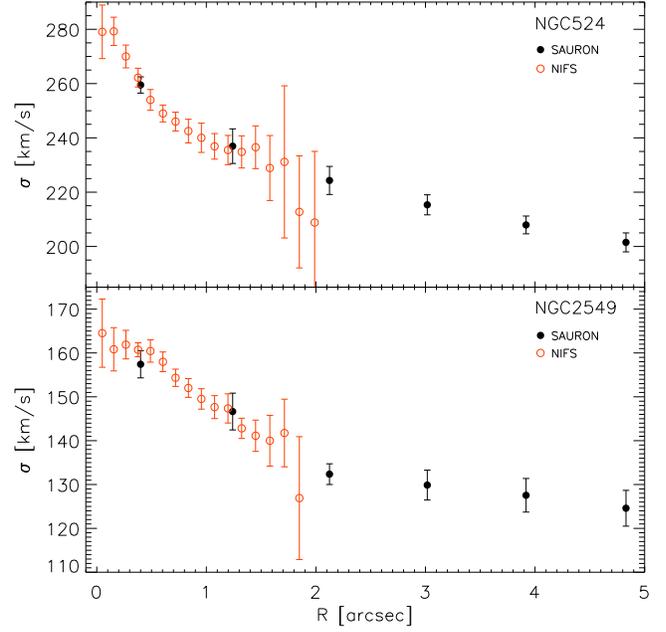}
\caption{\label{f:comparison} Comparison between NIFS (red open circles) and SAURON (filled circles) obtained by averaging data within concentric circular rings of different radii for NGC524 {(\bf top)} and NGC2549 {\bf(bottom)}. NIFS data for NGC524 were systematically lowered by 4\% to match the SAURON data.} 
\end{figure}

The SAURON kinematics was presented in \citet{2004MNRAS.352..721E}. Here we use an extraction which includes up to $h_6$ Gauss-Hermite moments and was done with Miles stellar  templates \citep{2006MNRAS.371..703S} in \citet{2007MNRAS.379..418C}. 

Figure~\ref{f:maps} presents two-dimensional kinematics extracted from our NIFS observations. The nuclei of both galaxies show significant rotation, with both rotation velocity and velocity dispersion being higher in the more massive NGC524. In both cases, $h_3$ and $V$ are strongly anticorrelated, while the $h_4$ maps are different: NGC524 has a clear dip in the central $0\farcs5$, while the $h_4$ map of NGC2549 is essentially constant and slightly positive over the whole field.

Since the NIFS kinematics probe a completely different spectral range than the SAURON observations, it is necessary to verify that these two data sets agree. The two sets of kinematics are widely separated in wavelength, and could therefore be expected to trace slightly different stellar populations with correspondingly different kinematics. Such population-dependent kinematics are not taken into account by our modelling process. Moreover, the absolute calibration of the velocity dispersion is sensitive to how well the intrinsic broadening of the template libraries (which for both data sets come from other instruments) mimics the instrumental dispersion. To obtain a reliable model with a meaningful fit, the two data sets must be mutually consistent within the assumed PSFs. In Fig.~\ref{f:comparison} we show a comparison between the velocity dispersions averaged on circular rings using bisquare weighting. The overlap is minimal; in radial direction NIFS data cover only about 2.5 SAURON pixels. Given the difference in the observational set ups and the uncertainties in the extraction of kinematics from such different spectral regions, it is remarkable to see such an agreement between the data sets. In the case of NGC524 we lowered the NIFS data for 4\% to obtain the match with the SAURON data, while this is, in principle, not necessary for NGC2549 NIFS data (a possible shift of 1\% would still be consistent with the uncertainties). The agreement between the velocity dispersion measurements is crucial for a robust determination of the mass of the black hole and, hence to constrain dynamical models of NGC524 we use the NIFS velocity dispersions systematically lowered for 4\% as shown in Fig.~\ref{f:comparison}.

%
%

\section{Determination of the Point Spread Function}
\label{s:res}

A robust estimate of the spatial resolution of our data is important for the construction of dynamical models, both in the sense of combining the data sets and in order to determine to which scales we can probe the internal dynamics of galaxies. Although the SAURON data were observed under natural seeing and the NIFS data were observed with LGS AO (see Section~\ref{ss:focus}) without any PSF reference stars in the field, we use the same general method in both cases to determine the actual PSF. The FWHM of the PSFs of the SAURON observations for NGC524 and NGC2549 are $1\farcs4$ and $1\farcs7$, respectively \citep{2004MNRAS.352..721E}, while NIFS PSFs derived below are listed in Table~\ref{t:psf}.

\begin{table}
   \caption{PSF of NIFS observations for NGC524 and NGC2549.}
   \label{t:psf}
$$
  \begin{array}{l|cc|cc}
    \hline
    \noalign{\smallskip}

   \multicolumn{1}{l|}{$galaxy$ } &   \multicolumn{2}{c|}{$NGC524$ } &  \multicolumn{2}{c}{$NGC2549$}\\ 

    $image$ & $HST$ & $MGE$ & $HST$ & $MGE$\\
    \noalign{\smallskip} \hline \hline \noalign{\smallskip}
  
    $fwhm$_{N} & 0.23 \pm 0.1 & 0.16 \pm 0.1  & 0.17 \pm 0.03 & 0.20 \pm 0.02  \\
    $fwhm$_{B} & 1.25 \pm 0.1 & 1.36  \pm 0.14   & 0.84 \pm 0.08 & 0.81 \pm 0.05  \\
    $inten$_{N} & 0.39 \pm 0.08 & 0.33 \pm 0.09  & 0.53 \pm 0.08 & 0.47 \pm 0.05\\ 
    \noalign{\smallskip}
    \hline
  \end{array}
$$ 
{Notes -- Table presents  parameters of two-Gaussian fits to HST and deconvolved MGE model images. In the first row there are values for FWHM (in arcseconds) of the narrow Gaussian, in the second row there are values for FWHM  (in arcseconds) of the broad Gaussian and in the third row is the intensity of the narrow Gaussian, where the sum of the intensities of the two Gaussians is one.}\looseness=-1
\end{table}

\begin{figure}
        \includegraphics[width=\columnwidth, bb=45 35 590 450]{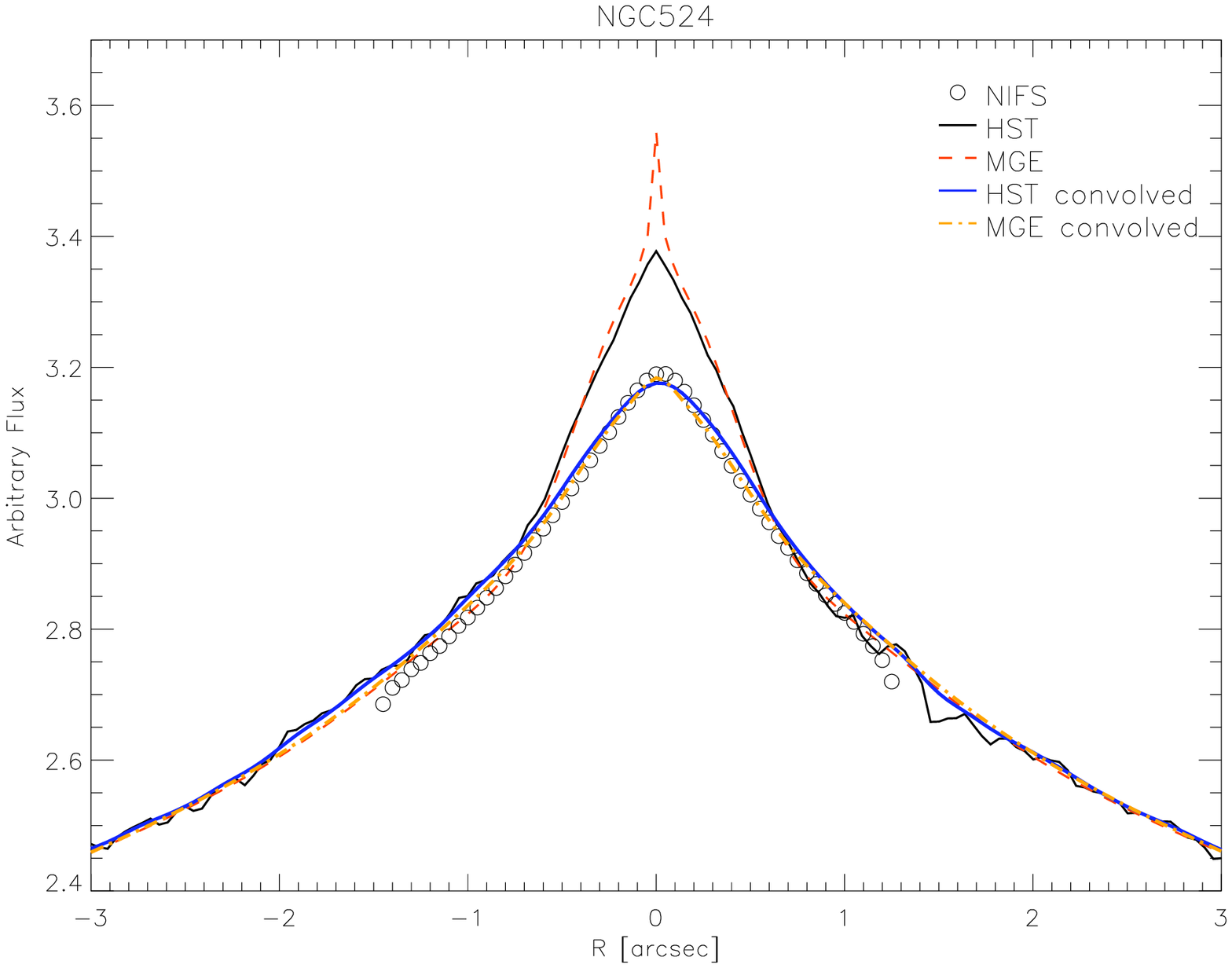}
        \includegraphics[width=\columnwidth, bb=45 35 590 450]{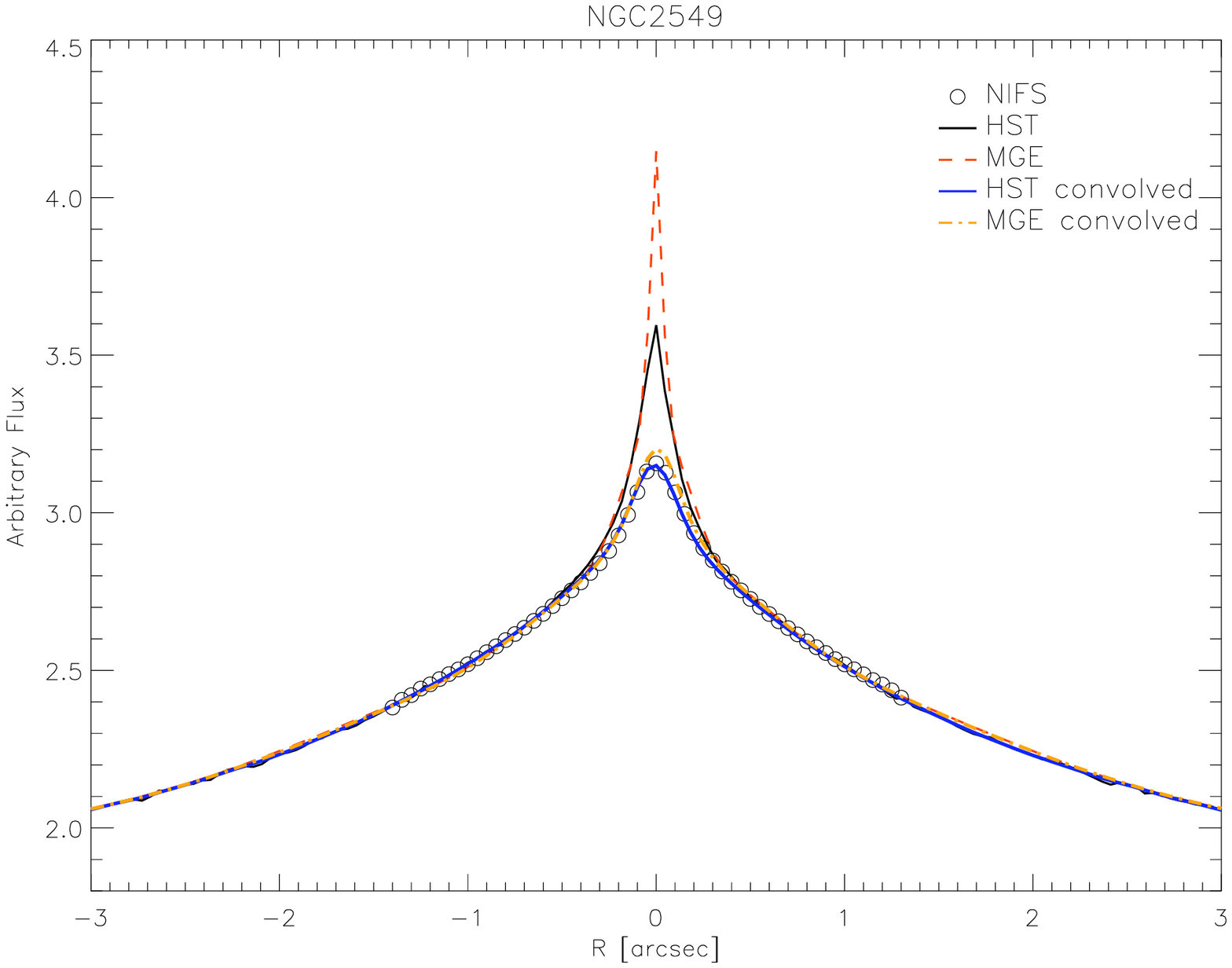}     
\caption{\label{f:PSFcomparison} Comparison of light profiles along the major axis between NIFS and HST images for NGC524 {\bf (top)} and NGC2549 {\bf (bottom)}. In both images NIFS profiles are shown with open symbols, HST profiles with black solid line, deconvolved MGE model profiles with red dashed line. HST and MGE profiles are convolved with appropriate PSFs from Table~\ref{t:psf}. Convolved HST and MGE profiles are presented by blue solid and orange dot-dashed lines.}
\end{figure}

Stars observed before and after the science frames show that there is minimal degradation of the AO correction during science observations (Section~\ref{ss:focus}) which could arise from incorrect adaptation of the system to changes in focus of the LGS. This, however, does not guarantee that the AO correction of science frames (and the final resolution of the galaxy nuclei) will be of sufficient quality. Moreover, an accurate description of the PSF is required to compare the observations with the model predictions. Since there are no point sources in the field-of-view (stars or AGN sources such as nuclei or jet knots), the only way to estimate the achieved PSF is to compare the reconstructed NIFS images with images of higher resolution: those obtained with the HST \citep[for a list of various methods to estimate the PSF see][]{2007astro.ph..3044D}. We first convolve the HST image with a concentric and circular double Gaussian description of the PSF, parameterised as the dispersions of the two components and their relative weight. The convolved image is then rebinned to the same pixel size as the NIFS observation. This image is compared with the reconstructed NIFS image, and the parameters of the PSF are varied until the best matching double-Gaussian is found. To obtain the final values one has to quadratically add the PSF of the HST image to the measured PSF. The double-Gaussian parameterization of the PSF is to some extent degenerate: different combination of the Gaussian parameters can reproduce the similar combined profiles. We estimate the uncertainties to the Gaussian parameters applying the method in different ways: we vary the initial conditions, we keep the centre of the images fix and let it vary freely, and we change the size of the NIFS maps limiting the effective area used in the comparison.  The same method was used previously for SAURON and OASIS data \citep{2004MNRAS.352..721E,2006MNRAS.373..906M,2006MNRAS.370..559S}. 

\begin{figure}
        \includegraphics[width=\columnwidth, bb=60 40 580 430]{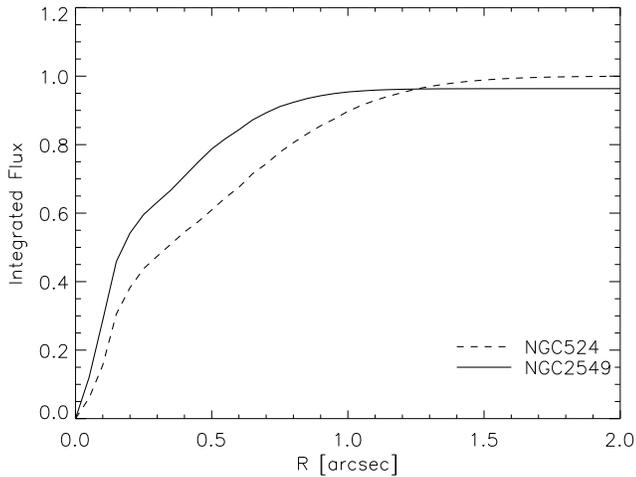}    
\caption{\label{f:intflux} Integrated flux from the determined PSF (using the HST images) for NGC524 (dashed line) and NGC2549 (solid line). The integrated flux is normalised such that the integral of the PSF for each galaxy is 1. }
\end{figure}

Implicit in this method is the assumption that the reference image (here coming from HST) has significantly better spatial resolution than the unknown image, such that the convolution kernel dominates the resulting image quality. The intrinsic spatial sampling of the HST/WFPC2 PC chip images is $0\farcs0455$ and the FWHM of the PSF of F702W and F814W images being used is $0\farcs07$ and $0\farcs08$, respectively, estimated using the TinyTim software \citep{TinyTim}. The final NIFS data-cubes have a pixel scale of $0\farcs05$ and in ideal conditions one could expect an AO-corrected PSF of $\sim0\farcs1$ FWHM as seen in Fig.~\ref{f:SFO}. These values are very similar to the HST/WFPC2 characteristics, and the assumption that the PSF of the reference image is negligible may not strictly hold. We continue following two approaches: one uses the HST/WFPC2 images directly, while the other uses a deconvolved model of the HST images as a PSF reference. In the latter case, we use the Multi-Gaussian Expansion (MGE) method \citep{1992A&A...253..366M,1994A&A...285..723E,2002MNRAS.333..400C} to parameterise the HST images and deconvolve them analytically by their respective PSFs. The MGE method and the same deconvolved models are also used in the construction of the mass model use as input to the dynamical models of our galaxies (see Section~\ref{ss:MGE} for more details).

The individual observations of NGC524 and NGC2549 differ in the quality of the AO correction. If one overplots isophotes of the reconstructed images, one can see that some images are more peaked in the centre while others have distorted  isophotes elongated in one direction. These differences arise from the changing LGS AO correction in response to the natural conditions. While one would ideally like to select only the best frames, there is a trade off between the achieved $S/N$ ratio and the size of the PSF.  Inspecting individual data cubes by eye, we selected those which had regular and concentric isophotes and merged them as described in Section~\ref{ss:merge}.  We verified that adding more lower quality frames worsens the measured PSF without significantly improving the $S/N$ ratio. The final PSFs of the data used in this work are listed in Table~\ref{t:psf}.

Figure~\ref{f:PSFcomparison} compares the PSF of our final NIFS data cubes for NGC524 and NGC2549 with the HST/WFPC2 and deconvolved MGE model images. They clearly show that our NIFS data are of lower spatial resolution than HST/WFPC2 images and, in this case, the MGE models are not really necessary for the PSF estimates, since the comparison with the deconvolved MGE model images yields very similar results to when using HST images directly. This is expected if the resolution of the data is sufficiently lower than the HST resolution in which case its PSF can be neglected. We therefore assume the PSF derived from the HST images directly, rather than introduce the added step of using the MGE model, and refer to these values for the rest of the paper.

For both galaxies, the narrow Gaussian is somewhat larger than the narrow component found with the tuning star (FWHM of $0\farcs23$ and $0\farcs17$ for NGC524 and NGC2549, respectively), while the poorer tip-tilt correction is most apparent in the larger width of the broad component (FWHM of $1\farcs25$ and $0\farcs81$ for NGC524 and NGC2549, respectively). NGC524 clearly has a poorer correction, but also uncertainties on the PSF components are larger, which is likely a consequence of core like light profile. As pointed out by \citet{2008Msngr.131....7D}, however, the Strehl ratio in LGS observations is not very dependent on the tip-tilt correction, since the flux within the core of the PSF does not change significantly if tip-tilt is present or not. We show the cumulative encircled energy of the PSF of our observations for NGC524 and NGC2549 in Fig.~\ref{f:intflux}. In both cases there is  approximately 40-50\% of total flux within $0\farcs2$.  The broad component of the PSF seems largely determined by the steepness of the galaxy's light profile, with the steeper profile of NGC2549 resulting in an overall narrower PSF and 90\% of the total flux within $0\farcs6$. For NGC524 90\% of flux is achieved only within about 1\arcsec.

%
%
\section{Dynamical models}
\label{s:dyn}

In this section we construct orbit superposition models defined by three integrals of motion and based on Schwarzschild's  method \citep{1979ApJ...232..236S, 1988ApJ...327...82R,1997ApJ...488..702R, 1998ApJ...493..613V}. Our models are axisymmetric and contain further developments which include fits to the two-dimensional stellar kinematics and are adapted for more general surface-brightness distributions. Due to their generality, axisymmetric three-integral models are standard for M$_{\bullet}$ determinations and exploration of the nuclear orbital structure in nearby galaxies \citep[e.g.][]{2003ApJ...583...92G, 2005ApJ...628..137V, 2007MNRAS.379..909N, 2008MNRAS.391.1629N, 2009ApJ...695.1577G}.   For more general triaxial models see \citet{2007MNRAS.376...71D,2008MNRAS.385..647V}. The implementation and adopted setup used here is described in \citet{2006MNRAS.366.1126C}. It consists of four steps: (i) the stellar light is parameterised, deprojected and converted to a stellar potential assuming axisymmetric shape and a stellar M/L; (ii) a representative (dithered) orbit library evenly sampling across the observable space (the three integrals of motion and covering the luminous mass of the galaxy) is constructed; (iii) orbits are projected onto the observable space (sky positions and LOSVD parameters) taking into account the PSF and apertures (Voronoi bins); (iv) weights to each orbit are determined using a non-negative least-squares fit \citep{1974slsp.book.....L}, which when co-added reproduce the observed stellar density and kinematics in each bin.

\subsection{The mass model}
\label{ss:MGE}

We used HST/WFPC2 F814W and F702W, and ground-based MDM {\it I} and {\it R} band images for NGC524 and NGC2549, respectively, to parameterise the surface brightness distributions of our galaxies with MGE models \citep{1994A&A...285..723E,2002MNRAS.333..400C}. Our MGE models were corrected for galactic extinction following \citet{1998ApJ...500..525S}, as given by NASA/IPAC Extragalactic Data base (NED) and converted to a surface density in solar units using the WFPC2 calibration from \citet{2000PASP..112.1397D}, while assuming for {\it I} and {\it R} band absolute magnitude for the Sun of 4.08 and 4.42 mag, respectively \citep{1998gaas.book.....B}. 

\begin{figure}
        \includegraphics[width=\columnwidth, bb=30 30 670 360]{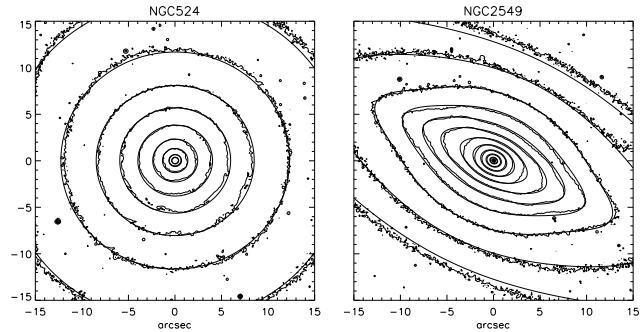}
\caption{\label{f:mge}HST WFPC2/PC images of NGC524 {\bf (left)} and NGC2549 {\bf (right)} with overplotted MGE models. Contours are spaced in half magnitude steps.}
\end{figure}

The MGE model for NGC524 was previously given in Table B1 of \citet{2006MNRAS.366.1126C}. We list the parameters of the MGE model for NGC2549 in Table~\ref{t:mge} of Appendix~\ref{s:mgepara} and plot the models over the HST images in Fig.~\ref{f:mge}. The minute differences between the MGE model and the image of NGC524 originate in the dusty central region of this galaxy. The galaxy isophotes are nearly round, and the MGE model was constructed to have axial ratios of all Gaussians equal to 0.95. NGC2549 is, however, an edge on galaxy and with a peculiar isophotal shape. While it is usually classified as a normal S0 \citep[e.g.][]{1991trcb.book.....D}, this object has several characteristics which indicate a possible bar: its isophotes change from being strongly boxy ($0\farcs7-4\arcsec5$) to strongly disky ($4-20\arcsec$ and beyond) \citep{2001AJ....121.2431R}, an 'X' shape distribution of light in the bulge (giving rise to strongly boxy isophotes) visible on HST images, and evidence for a ring-like structure at larger radii (visible on ground based colour images). The variation from strongly boxy to disky isophotes is difficult to reproduce with only positive Gaussians, and hence the largest discrepancies (a few percent) in our MGE model arise in that region ($4-5\arcsec$).

We assume that both galaxies can be well reproduced by an axisymmetric shape. These galaxies do not show isophotal twists and their kinematics have clear disk-like rotation \citep{2008MNRAS.390...93K}.  Since a proper dynamical treatment of bars is beyond the scope of this paper, we keep our axisymmetric assumption as the next closest representation of NGC2549. The bias introduced by modelling a likely barred galaxy using a model with a static, axisymmetric potential has so far not been well explored. What is clear, however, is that as we probe systems with lower black hole masses, bars become a common feature of the general dynamical make-up of the hosts. The degree to which the large-scale dynamical structure can affect the central dynamics is difficult to estimate, and as we shall discuss in Section~\ref{s:discuss}, characterising the large-scale galaxy can help constrain M$_{\bullet}$. In the current absence of a better model, however, we will proceed as others have done using a static, axisymmetric model \citep[e.g.][]{2007ApJ...670..105O,2009ApJ...693..946S}.

Deprojection of the surface brightness of an ellipsoidal body is formally non-unique for all but the edge-on case \citep{1987IAUS..127..397R,1988MNRAS.231..285F}. In practice, determination of the viewing angles through stellar dynamical models is also found to be degenerate  \citep{2005MNRAS.357.1113K,2008arXiv0811.3474V}. It is only possible to obtain an estimate of the inclination for such objects by applying further assumptions, such as of the shape of the velocity ellipsoid \citep{2008MNRAS.390...71C}, using the observationally motivated positive anisotropy \citep{2007MNRAS.379..418C}. The accuracy of M$_{\bullet}$ determinations from stellar dynamics are dependent on these effects, since one must choose an inclination to fix the potential. For the two galaxies considered here, there are reliable indications of their true inclinations. NGC524 has a large dust disk, which, if assumed to be intrinsically circular, gives an inclination of $i \sim 20\degr$ \citep{2006MNRAS.366.1126C}. NGC2549 is close to edge on, which is visible both from photometry and kinematics \citep{2008MNRAS.390...93K}.

While there are difference in stellar populations between NGC524 and NGC2549, within each galaxy, the variation line-strength indices are mild across the SAURON field \citep{2006MNRAS.369..497K}. Kuntschner et al. (2009, in prep.) derive the age (single stellar population equivalent) within Re/8 apertures to be old ($>12$ Gyr) and intermediate (5.5 Gyr) in NGC524 and NGC2549, respectively. Since M/L ratio is mostly a function of the $H_{\beta}$ line-strength index and, hence, predominantly the age of stellar population \citep[see Fig.~16 in][]{2006MNRAS.366.1126C}, which changes mildly across the fields, we assume in our dynamical models a constant M/L. We also assume that the dark matter contributes only with a small fraction in the central regions \citep{2001AJ....121.1936G, 2003ApJ...595...29R, 2006MNRAS.366.1126C, 2006ApJ...649..599K, 2007MNRAS.382..657T, 2008ApJ...684..248B} and do not explicitly add it to the models.

Finally, in this paper we use distances from \citet{2001ApJ...546..681T}, adjusted for the Cepheid zero-point of \citet{2001ApJ...553...47F} putting NGC524 and NGC2549 at 23.3 and 12.3 Mpc, respectively. 

\begin{table}
   \caption{\bf Summary of relevant properties for NGC524 and NGC2549}
   \label{t:stars}
$$
  \begin{array}{l|l|l|c}
   \hline
    \noalign{\smallskip}

$Galaxy property$& $NGC524$ &$NGC2549$& source\\
    \noalign{\smallskip} \hline
      $Morphological type$         & S0^{+}(rs) & S0^{0}(r)sp  &1\\
      $Nuker $ \gamma                   & 0.03 & 0.67 			& 2\\
      $Nuclear type$                       & core & cusp			 & 2\\
      $Effective radius  [arcsec]$  & 51    & 20      			& 3\\
      \sigma_e $[km/s]$                & 235 & 145    			&3\\
      $Distance  [Mpc] $                & 23.3  & 12.3    	        	&4\\
      $Inclination [degrees]$         & 20      &  90      		&5\\
      \noalign{\smallskip}
    \hline
  \end{array}
$$ 
Notes -- Sources (if two references present, they refer to galaxies respectively): 1 - \citet{2004MNRAS.352..721E};  2 - \citet{2007ApJ...664..226L, 2001AJ....121.2431R}; 3 - \citet{2007MNRAS.379..401E}; 4 - \citet{2001ApJ...546..681T}; 2 - \citet{2006MNRAS.366.1126C}. 
\end{table}

\begin{figure*}
        \includegraphics[width=\columnwidth, bb=30 30 450 590]{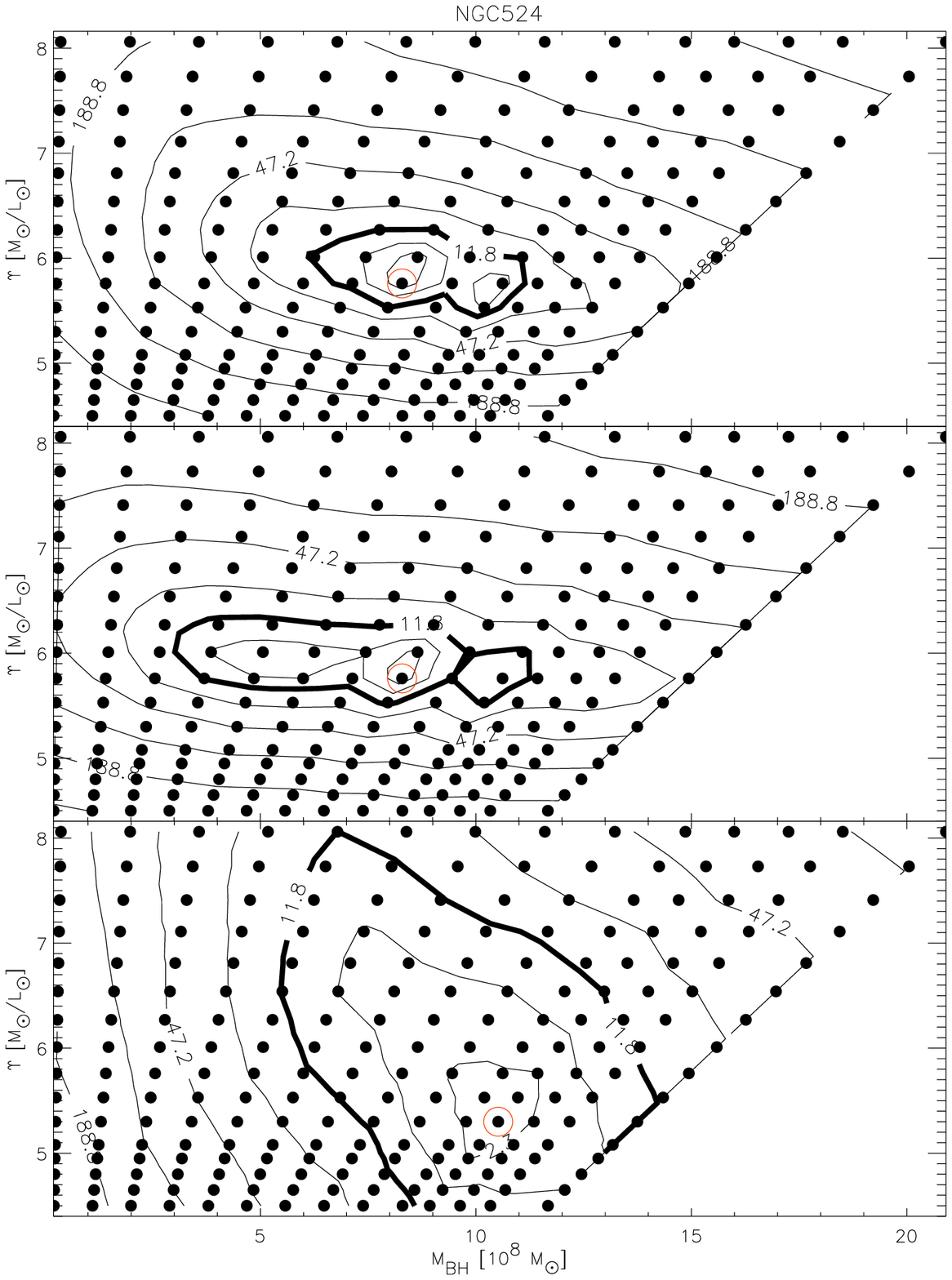}
        \includegraphics[width=\columnwidth, bb=30 30 450 590]{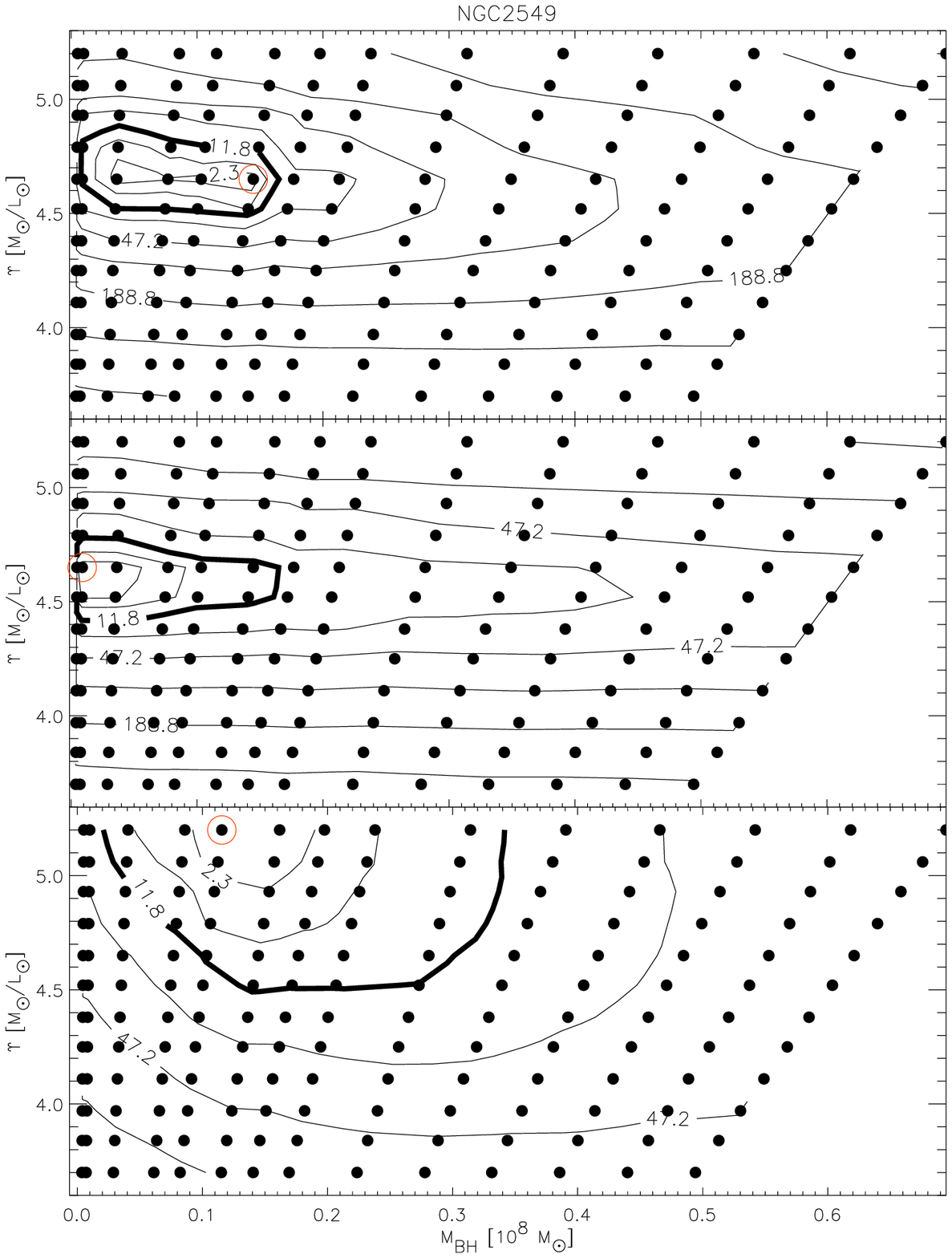}     
\caption{\label{f:grids} Schwarzschild dynamical models and the determination of the best fitting parameters for NGC524 {\bf(left)} and NGC2549 {\bf(right)}. Each symbol is a dynamical model specified by given M/L and M$_{\bullet}$. The agreement between the data and the models are described by overploted contours $\Delta  \chi^2 = \chi^2 - \chi^2_{min}$ contours showing 1, 2 and 3$\sigma$ levels for two parameters. Further contours are spaced by a factor of 2. Open (red) circles mark the best fitting models. {\bf Top} panels show grids of models constrained by both SAURON and NIFS kinematics, {\bf middle} panels show grids of models constrained by SAURON only kinematics and {\bf bottom} panels show models constrained only using NIFS data.}
\end{figure*}

\subsection{Stellar dynamical models}
\label{ss:stdyn}

Our Schwarzschild models were simultaneously fitted to the SAURON and NIFS kinematics and mass distributions parameterised by the MGE models. Since the models are axisymmetric and by construction point-(anti)symmetric, we symmetrised the kinematic maps using kinematic position angles (PA$_{kin}=1\degr$ and $40\degr$ for NGC524 and NGC2549, respectively) determined in \citet{2007MNRAS.379..418C}.  The symmetrisation uses the mirror-(anti)-symmetry of the kinematic maps, such that kinematic values from for positions ((x,y,), (x,-y), (-x,y), (-x,-y)) were averaged. Since the Voronoi bins have irregular shapes and they are not equally distributed with respect to the symmetry axes, we average the four symmetric points, and, if for a given bin there are no bins on the symmetric positions, we interpolate the values on those positions and average them. We keep the original errors of these bins in order not to underestimate the parameter uncertainties. The assumption here is that the uncertainties of the kinematic parameters in bins at positions (x,y,), (x,-y), (-x,y), (-x,-y) are similar, which is reasonable given that the light falls off symmetrically in all four quadrants. We exclude SAURON bins in the central $1\farcs3$  (9 bins), since these overlap with the high resolution NIFS observations. We constrain the dynamical models fitting the kinematics by up to $h_6$ Gauss-Hermite moments for both data sets.

Each dynamical model is defined by two free parameters: M/L and M$_{\bullet}$. The orbit library is constructed for a given M$_{\bullet}$ at a given (expected) M/L and consists of 444528 orbits, which are bundled in groups of $6^3=216$ before the linear orbital superposition. It is, however, not necessary to compute the orbit library for each pair of (M/L, M$_{\bullet}$). The predictions of the orbit library can be scaled to construct models for different M/L ratios. We use a modest amount of regularisation $\Delta=10$ \citep[as defined in][]{1998ApJ...493..613V}. Figure~\ref{f:grids} shows the main results of this study: for a set of  (M/L, M$_{\bullet}$) pairs defining individual models overploted are contours of $\chi^2$, showing the agreement between the kinematic data and the model predictions.

In the case of NGC524, the best fitting model has  M$_{\bullet}$=($8.3^{+2.7}_{-1.3} )\times 10^8$ M$_{\sun}$ and M/L $= 5.8 \pm 0.4$ ({\it I} band). Note that the M/L estimate for the almost face-on NGC524 is strongly dependent on the assumed inclination \citep[see Figure A1 in ][]{2006MNRAS.366.1126C}.  For NGC2549, the best fitting model is the one with  M$_{\bullet}$=($1.4^{+0.2}_{-1.3} )\times 10^7$ M$_{\sun}$  and M/L $= 4.7 \pm 0.2$ ({\it R} band). The quoted uncertainties for M$_\bullet$ are $3\sigma$ values marginalised over the M/L and correspond to one degree-of-freedom ($\Delta \chi^2=9$). 
Due to numerical uncertainties inherent to Schwarzschild's method we advise using 3$\sigma$ errors only and we do not measure the uncertainties at 1 and 2 $\sigma$ levels \citep[for a discussion on topology of $\chi^2$ contours see Appendix A of][]{1998ApJ...493..613V}. Since errors are expected to scale as $\sqrt{\Delta \chi^2}$ \citep{1992nrfa.book.....P} one can estimate 1 and 2 $\sigma$ confidence levels as approximately 1/3 and 2/3 of the $3\sigma$ values, respectively. Similar approach was used by \citet{2003ApJ...583...92G} to estimate 1 $\sigma$ confidence levels of studies that only provide 3$\sigma$. In addition to models fitted to both the SAURON and NIFS data, we also constructed models constrained with only NIFS and only SAURON data, in order to asses the relative importance of these data sets. This exercise demonstrates that the best-fitting parameters are consistent within $3\sigma$ level between models constrained fitting different data sets. 

More importantly, this demonstrates that the SAURON data, whilst being relatively insensitive to the black hole mass due to its lower spatial resolution, provides a critical constraint on the M/L by virtue of its large-scale spatial coverage.  The M/L and M$_{\bullet}$ are not independent, and the drastic reduction in uncertainty of the M/L provides a significant reduction in the uncertainty of M$_{\bullet}$. Without the SAURON data, the uncertainty for M$_\bullet$ doubles, while errors on M/L increase between 5 to 10 times. The NIFS data are however crucial to determine M$_{\bullet}$, especially for low mass M$_{\bullet}$ when $R_{sph}$ becomes much smaller than a SAURON resolution element; 'SAURON only' model for NGC2549 can only give an upper limit to M$_{\bullet}$, and NIFS data are necessary to determine it robustly \citep[see also][]{2006MNRAS.370..559S}. 
 
\begin{figure*}
        \includegraphics[width=\textwidth, bb=30 30 990 220]{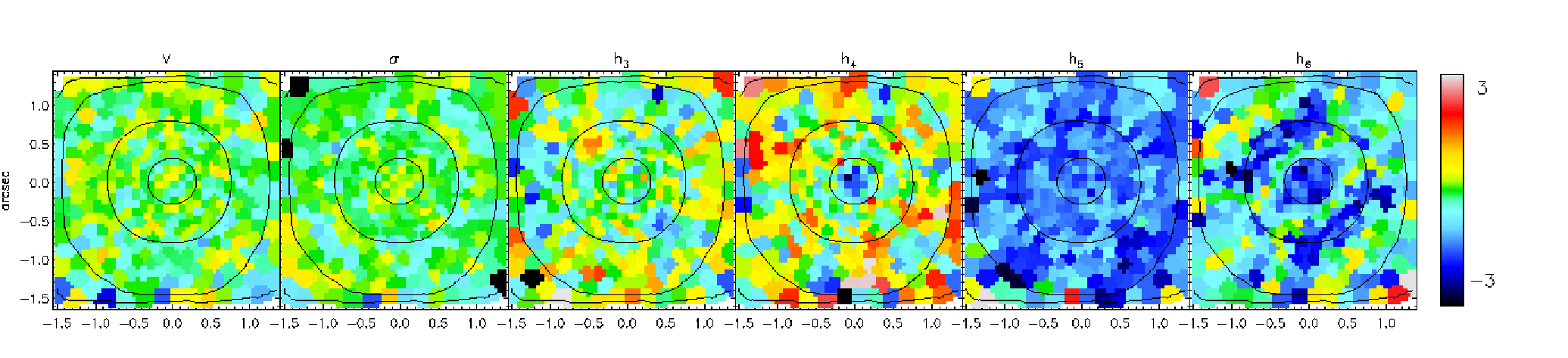}
        \includegraphics[width=\textwidth, bb=30 30 1115 210]{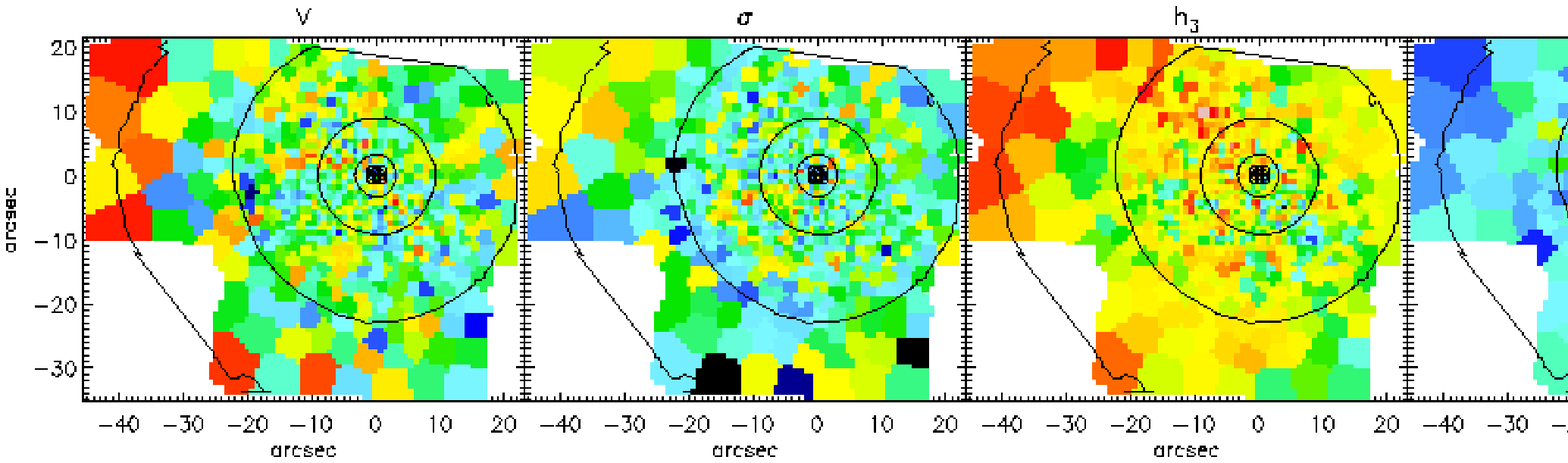}     
\caption{\label{f:modelmaps1} Residuals between the data and the best fit Schwarzschild model for NGC524 presented by NIFS {\bf (top)} and SAURON {\bf (bottom)} data. The residuals were obtained subtracting the model from the data and dividing by the errors. From left to right the columns show: mean velocity $V$, velocity dispersion $\sigma$, and Gauss-Hermite moments $h_3 - h_6$. Overplotted contours are surface brightness isophotes. Circles in the center of the SAURON maps show bins excluded from the fit. North is up and East to the left. SAURON kinematic maps can be found in \citet{2004MNRAS.352..721E}. }
\end{figure*}

\begin{figure*}
        \includegraphics[width=\textwidth, bb=30 30 990 220]{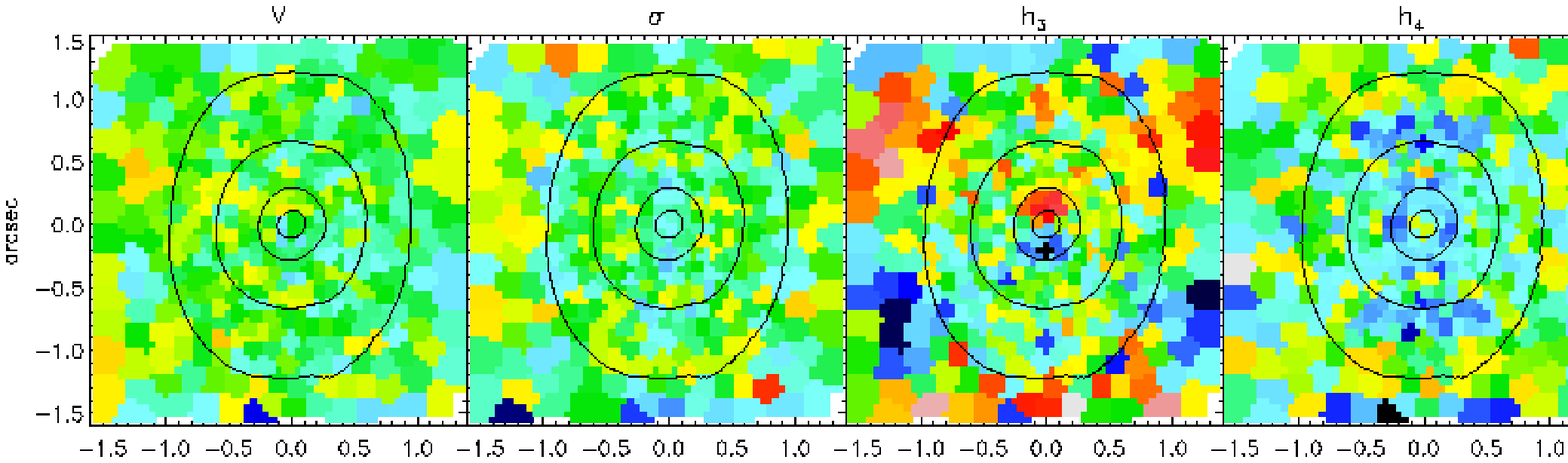}     
        \includegraphics[width=\textwidth, bb=30 30 990 250]{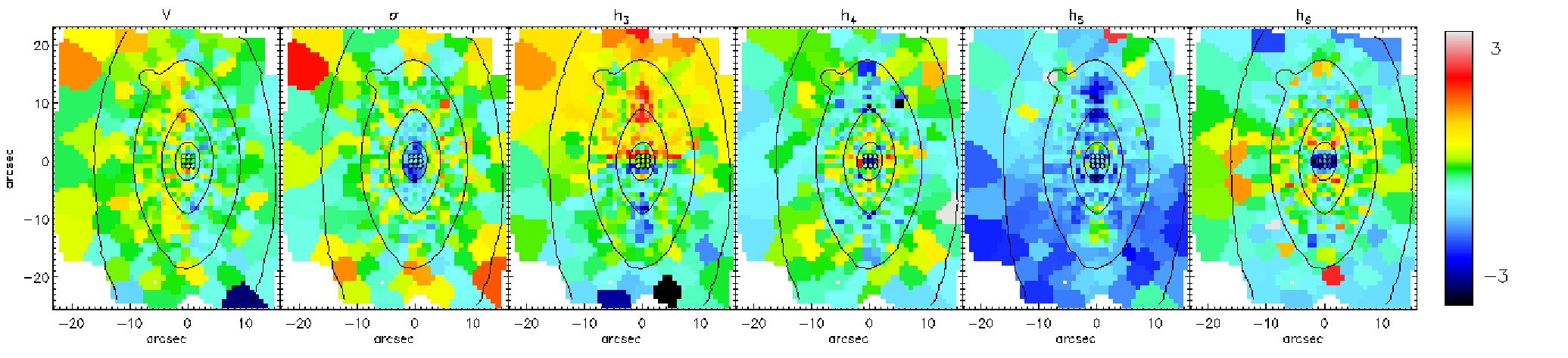}     
\caption{\label{f:modelmaps2}  Residuals between the data and the best fit Schwarzschild model for NGC2549 presented by NIFS {\bf (top)} and SAURON {\bf (bottom)} data. The residuals were obtained subtracting the model from the data and dividing by the errors. From left to right the columns show: mean velocity $V$, velocity dispersion $\sigma$, and Gauss-Hermite moments $h_3 - h_6$. Overplotted contours are surface brightness isophotes. Circles in the center of the SAURON maps show bins excluded from the fit. North is up and East to the left. SAURON kinematic maps can be found in \citet{2004MNRAS.352..721E}. } 
\end{figure*}

In Figs.~\ref{f:modelmaps1} and~\ref{f:modelmaps2} we show the residuals between the data and the best fit models for NGC524 and NGC2549, respectively. The residuals were calculated by subtracting the predictions of the best fit model from all kinematic constraints ($V$, $\sigma$, $h_3$ - $h_6$) and dividing by the associated errors. The stretch in colour is chosen to show the goodness of fit relative to the kinematic errors: the extremes bound 3 times the uncertainty. Most of bins, for both NGC524 and NGC2549, have green, yellow or light blue colours which mean that the residuals are within the kinematic errors. Somewhat higher residuals ($\geq 2\sigma$ level) are systematically visible only for higher Gauss-Hermite moments ($h_5$, $h_6$).

\begin{figure}
        \includegraphics[width=\columnwidth, bb=30 30 590 590]{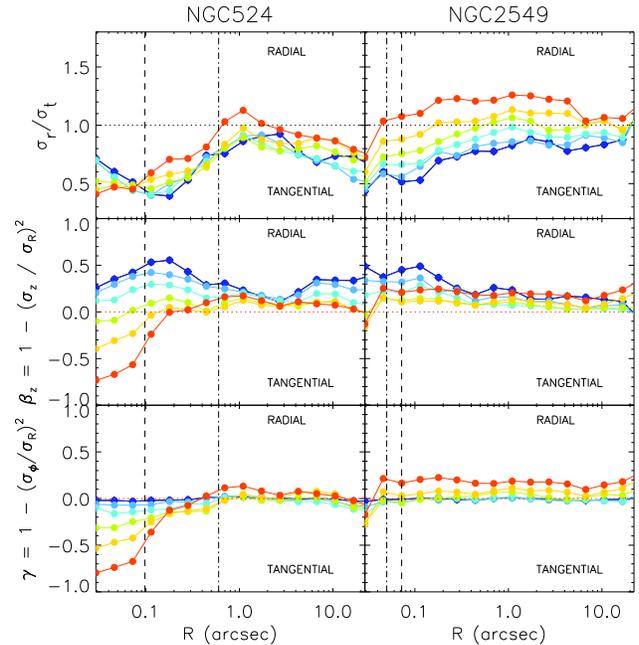}
\caption{\label{f:aniso}  Radial profiles of anistropy of Schwarzschild dynamical models for NGC524 {\bf (left)} an NGC2549 {\bf (right)}. {\bf Top:} anisotropy parameterised in spherical coordinate system by $\sigma_r/\sigma_t$. {\bf Middle:} anisotropy parameterised in cylindrical coordinate system by $\beta_z$. {\bf Bottom:} anisotropy parameterised in cylindrical coordinate system by $\gamma$ (see text for definitions).  Anisotropy was measured at different polar angels defined in the meridional plane, ranging from $15\degr$ (close to the equatorial plane, red line) to $75\degr$ (close to the symmetry axis, blue line). The change in colours shows the increase in the angle from the major to minor axis (red to blue colours). Dashed vertical lines show $\sigma$ of the narrow component of the LGS PSF and dot-dashed vertical lines show R$_{sph}$ using the derived M$_\bullet$. }
\end{figure}

The general nature of the three integral models allows us to investigate the orbital anisotropy of our best fit models. We do this by plotting different coordinates of the velocity dispersion tensor. We first define the ratio of  radial to tangential velocity dispersions, $\sigma_r/\sigma_t$ in Fig.~\ref{f:aniso}, where $\sigma_t^2=(\sigma^2_\theta + \sigma^2_\phi)/2$, in spherical coordinates ($r$, $\theta$ and $\phi$). Since $\sigma_\phi$ includes only random motion, an isotropic system will have $\sigma_r/\sigma_t=1$.  This measure of anisotropy has been used in some previous studies of nuclear orbital structure \citep{2003ApJ...583...92G, 2006MNRAS.370..559S, 2006MNRAS.367....2H, 2009ApJ...695.1577G}, but we note that it is mostly suited for description of spherically symmetric objects. NGC2549 is an edge on galaxy and clearly not spherical, while the low inclination and the projected flattening of MGE Gaussians determine the intrinsic flattening of NGC524 to be $q_i\sim0.41$ \citep[eq.~(9) of][]{2002MNRAS.333..400C}, and again not a spherical galaxy\footnote{The regions under the direct influence of SMBHs tend to be spherically symmetric allowing a representation of the moments of the velocity ellipsoid in spherical coordinates.}. For these reasons we also make use of the  components of the velocity dispersion tensor in cylindrical coordinates ($R$, $\phi$ and $z$): $\sigma_R$, $\sigma_\phi$ and $\sigma_z$. We define anisotropy parameters $\beta_z=1 - (\sigma_z/\sigma_R)^2$ and $\gamma = 1 - ( \sigma_\phi/\sigma_R)^2$. The first parameter, $\beta_z$, describes the shape of the velocity ellipsoid in ($v_R$, $v_z$) plane, a plane including the symmetry axis plane of an axisymmetric galaxy, where $\beta_z=0$ if the cross-section of the velocity ellipsoid is a circle. The second parameter, $\gamma$, describes the shape of the velocity ellipsoid in ($v_R$, $v_\phi$), a plane orthogonal to $v_z$, which can be identified with the equatorial plane in an axisymmetric galaxy. Again, if $\gamma = 0$, the shape of the velocity ellipsoid in this plane is a circle. Departures from the circular shape of the velocity ellipsoid quantify the bias towards radial ( $\beta_z> 0$, $\gamma> 0$) or tangentially  ( $\beta_z<0$, $\gamma <0$) biased orbital distributions.

In Fig.~\ref{f:aniso} we plot these three anisotropy parameters for NGC524 and NGC2549 as as function of radius and position angle within each galaxy. The galaxies show similar orbital structures at large radii, but markedly different orbital structure close to the black holes. In the spherical coordinates, orbits in NGC524 are clearly tangentially anisotropic. The significant drop in $\sigma_r/\sigma_\phi$ occurs at the radius similar to the $R_{sph}$, which also corresponds well to the radius within which there is a significant drop in measured values of the $h_4$ Gauss-Hermite moments. To see this effect in cylindrical coordinates one has to keep in mind the definition of the coordinate systems, where the $r$ and $R$ coordinates have the same orientation on the major axis, but a perpendicular on the minor axis, hence the difference between the minor and major axis in $\beta_z$ parameters. The equatorial plane view of $\gamma$ anisotropy values are also consistent whit this picture. The orbital bias is tangential, even without the streaming motion present in anisotropy parameters. Note that for the minor axis $\gamma$=0 by symmetry for an axisymmetric galaxy. For NGC524 we can conclude that both spherical and cylindrical coordinate systems give a clear picture of orbits being mostly tangentially biased in the central $0\farcs5$. At larger radii, where the flatness of the galaxy is more dominant, the anisotropy parameters defined in the cylindrical coordinates shows that $\sigma_z < \sigma_R$, but  $\sigma_R \approx \sigma_\phi$, which means that the velocity ellipsoid in NGC524 is very close to oblate \citep[as shown in][]{2007MNRAS.379..418C}.
 
NGC2549 shows a somewhat different picture in the nucleus. The spread of anisotropy in the $\sigma_r/\sigma_t$ is inconclusive regardless the anisotropy (or possibly weakly tangentially anisotropic). On the other hand, $\beta_z$ clearly shows that anisotropy is very similar along both major and minor axis, and in generally $\sigma_z < \sigma_R$, without a notable change towards the central black hole, as it is visible for NGC524. A possibly important difference is that for NGC524 the resolution of our observations is several times less than the R$_{sph}$, while in NGC2549 they are only comparable, and a possible tangential anisotropy structure around the SMBH is not visible. In the equatorial plane, $\gamma$ values along most position angles are basically consistent with zero. The values of $\beta_z$ and $\gamma$ imply that NGC2549 has oblate velocity ellipsoid. The anisotropy of NGC2549 over the SAURON field of view is in an excellent agreement with the results from Jeans modelling with constant anisotropy \citep[$\beta_z=0.17$ and $\gamma=0$, Fig.~5 in][]{2008MNRAS.390...71C}.

%
%
\section{Discussion}
\label{s:discuss}

\subsection{Relative importance of NIFS and SAURON data}
\label{ss:imp}

The comparison between the grids of models constrained by NIFS and SAURON, only NIFS and only SAURON data (Fig.~\ref{f:grids}) show the importance of combined large scale and high resolutions data sets. Stellar dynamical models constrained by high-resolution data with small field-of-view are not sufficient to robustly determine M$_\bullet$. The main reason is that they are not able to constrain the orbital distribution and hence the total M/L of the galaxy. This reflects the need to establish the importance of stars on radial orbits, which may pass close to the SMBH, whilst spending most of their time at much larger radii. In the lower panels of Fig.~\ref{f:grids} the models constrained by NIFS data alone show a clear degeneracy between the M/L and M$_\bullet$ where low M/L is increasing M$_\bullet$ and vice versa. While the formal best fitting M$_\bullet$ are not very different from the one determined using both NIFS and SAURON data, the uncertainties are much larger: a factor of a few in M$_\bullet$ to about an order of magnitude increase in M/L uncertainties. The SAURON data (middle panel on Fig.~\ref{f:grids}) constrain the M/L of our galaxies, because they cover more than 1 $R_e$, which is consistent with finding of \citet{2005MNRAS.357.1113K} that large scale data carrying information on the orbital structure over approximately 1$R_e$ are necessary to constrain the distribution function. The implications are clear: to accurately measure the central orbital structure and gravitational potential (i.e. including M$_\bullet$) in galaxy nuclei requires IFU kinematics on both small and large scales. Similar results, using high resolution and large scale IFU data, were reported in \citet{2004A&A...415..889C} and \citet{2006MNRAS.370..559S}, except that in the case of NGC2549, a smaller galaxy compared to the galaxies in these studies, the high resolution data are actually crucial to determine the lower limit to M$_\bullet$. 

The sphere of influence for NGC524 is $R_{sph} \sim 0\farcs6$ and for NGC2549 $R_{sph} \sim 0\farcs05$ using the masses derived in this study. In the case of NGC524 the  $R_{sph}$ is well resolved by the LGS observations, while the SAURON data have the sigma of the Gaussian PSF quite similar ($0\farcs6$). In the case of NGC2549, the sigma of the narrow Gaussian of the LGS observations ($0\farcs07$) is similarly just larger than the R$_{sph}$. While neither SAURON data for NGC524 nor NIFS data for NGC2549 formally resolve respective R$_{sph}$, they do constrain the M$_\bullet$. The reason likely lies in the combination of high S/N of data, IFU coverage  of the nuclear region and a significant part of the galaxy and that sigmas of the PSFs are similar to the $R_{sph}$, partially resolving the spheres of influence. This finding is important for future dynamical studies of SMBHs as well as for studies of M$_\bullet - \sigma_e$ relation. A discussion of the biasses introduced when M$_\bullet$ measurements based on data which do not fully resolve R$_{sph}$ are excluded from determining the properties of M$_\bullet - \sigma_e$ relation, is given in \citet{2009ApJ...698..198G}. We, however, note that likely the crucial contribution for determination of M$_\bullet$ in marginally resolved cases comes from two-dimensional coverage of the kinematic constraints. 

\begin{figure*}
        \includegraphics[width=\textwidth, bb=20 60 1010 335]{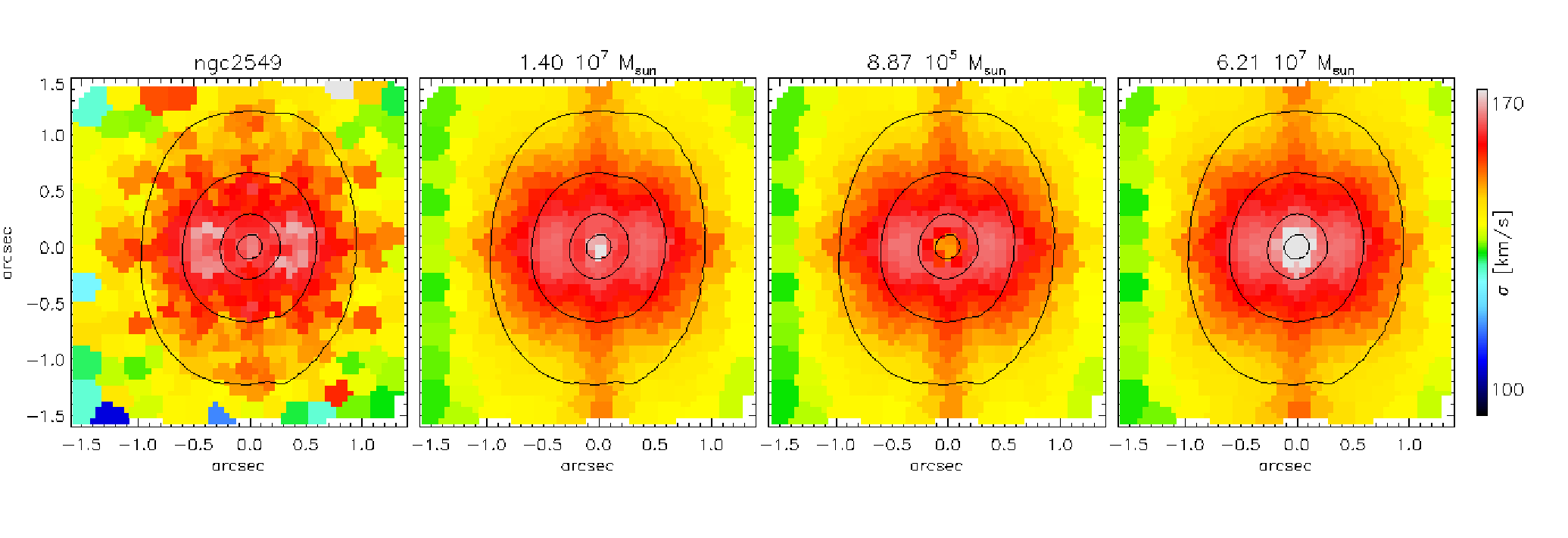}
        \includegraphics[width=\textwidth, bb=20 60 1010 335]{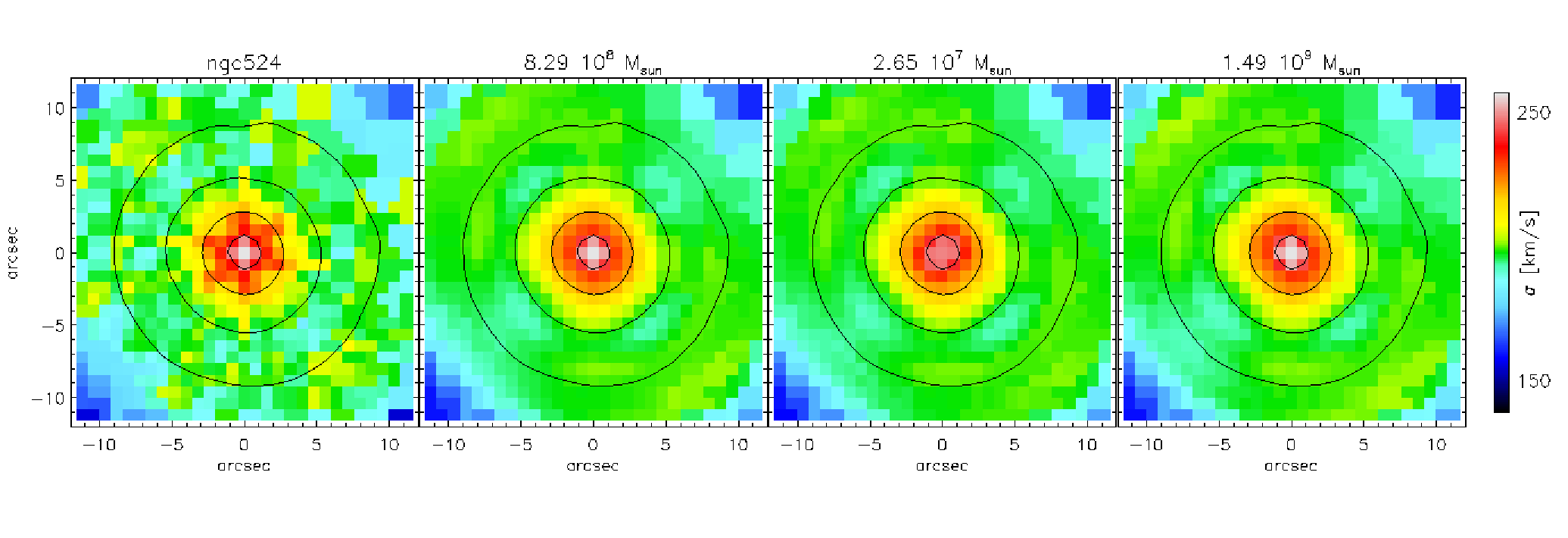}
\caption{\label{f:sigmap} Comparison between data and model velocity dispersion maps for {\bf (top)} NGC2549 and NGC524 {\bf (bottom)}. From left to right: $\sigma$ (symmetrised) and model prediction maps for different M$_{\bullet}$ (at the best fitting M/L). The best fitting model is shown in the map adjacent to the observed $\sigma$ map, followed by a model with a too small M$_{\bullet}$ and a model with too large M$_{\bullet}$, respectively. The models shown here for NGC2549 were constrained with both SAURON and NIFS data, while the models shown here for NGC524 were constrained using only SAURON data. Note the colour change in the central few pixles which are a consequence of change in M$_{\bullet}$. North is up and east to the left. }
\end{figure*}

In order to illustrate that our models can robustly measure M$_\bullet$ in these galaxies, in spite of marginally resolving R$_{sph}$,  in Fig.~\ref{f:sigmap}  we show clearly visible difference between the data and various models. Here we concentrate on the models constrained by SAURON and NIFS data for NGC2549 and only SAURON data for NGC524. In the latter case we want to stress that even SAURON data, without the help of high spatial resolution data are able to distinguish between models with too large and too small M$_\bullet$. In the upper panels of the Fig.~\ref{f:sigmap} we show the difference between velocity dispersion maps for three models (SAURON + NIFS constraints) at the best fitting M/L for different M$_{\bullet}$ in NGC2549: the best fitting model, a model with too small and a model with too large M$_{\bullet}$. The difference between models is mostly evident in the central $\sigma$ peak: a too low M$_\bullet$ underpredicts, while a too high M$_\bullet$ overpredicts it. The contribution of the NIFS data in constraining the lower limit on the M$_{\bullet}$ in NGC2549 is not only statistical and visible in $\chi^2$ values, but can also be judged by eye. In the lower panels of Fig.~\ref{f:sigmap} we plot a comparison between the velocity dispersion maps for data and three models (SAURON only constraints): the best fitting model, a model with too small and a model with too large M$_{\bullet}$. Again, the only relevant change between the model maps occurs in the central few pixels, with the largest change in the central pixel, confirming that even using only SAURON data can constrain the M$_{\bullet}$ in NGC524, although the uncertainties are larger than when NIFS data are added.

\subsection{Accuracy of M$_{\bullet}$ determination on the PSF}
\label{ss:psfsim}

Observing nearby galaxies with AO systems is not practical unless it is possible to use the LGS without relaying on nearby tip-tilt stars, such as in this work. The correction of the PSF, however, in similar cases will depend on the properties of the central light profiles. From that point of view we want to establish to what accuracy it is possible to measure M$_\bullet$ covering a range of corrections achievable with current LGS systems.

We tested this by constructing two simple model galaxies described by a de Vaucouleurs R$^{1/4}$ law, of total mass $5\times10^{10}$ M$_\odot$ and size $R_e=27\arcsec$, also adding an SMBH of 2\% and 1\% of the galaxy mass in each of them. When put at a distance of Virgo (16.5 Mpc), the SMBH in these model galaxies had estimated R$_{sph}$ of $0\farcs27$ and $0\farcs13$, respectively.   We assumed a spherical symmetry and used Jeans models \citep[the JAM implementation of][]{2008MNRAS.390...71C} to predict the second kinematic moments as a function of radius, given the light of our model galaxies parametrized with the MGE method and a realistic PSF. We made a 100 different model predictions of second moments (V$_{rms}$), each with a different PSF. The PSF was parameterised as a double Gaussian where the FWHM of the broad component is kept fixed to a (natural) seeing of $0\farcs8$, while the narrow component model was varied from $0\farcs04$ (approximating the diffraction limit at 8-10 meter telescopes in the near-infrared) to $0\farcs5$. The relative flux ratio of the components was such that their sum is equal to unity and we varied the flux of the narrow component from 0.1 to 0.5. 

\begin{figure}
        \includegraphics[width=\columnwidth, bb=30 30 450 590]{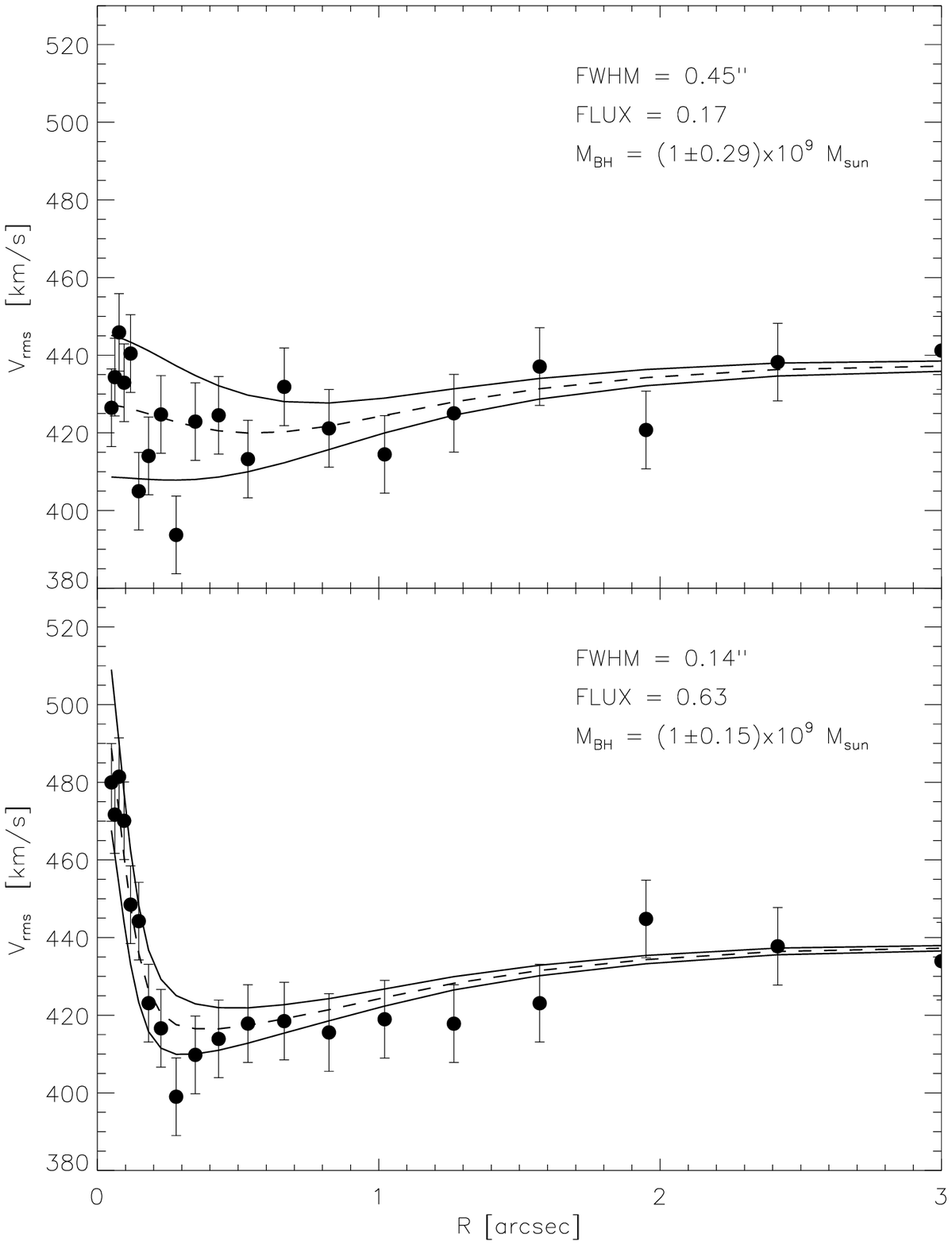}
\caption{\label{f:psfplot} Two examples of the simulations used to test the influence of the PSF on the accuracy of the M$_\bullet$ estimates. {\bf Top:} Predictions for model with a poor PSF given by FWHM and the flux of the narrow component of 0\farcs45 and 0.17, respectively. {\bf Botom:} Predictions for model with an excellent PSF given by FWHM and the flux of the narrow component of 0\farcs14 and 0.63, respectively. In both panels, the symbols represent the V$_{rms}$ of the galaxy model which are fit with Jeans models. The best fit model is given by the dashed line, and the two lines above and below the best fit model are models that are at $\Delta \chi^2=9$ level away from the best fit $M_\bullet$}
\end{figure}

We then added a constant error of 10 km/s to the derived V$_{rms}$ and fitted them with the spherical Jeans models, but now varying the M$_\bullet$ (100 different values centred on the true value) and accounting for M/L by scaling the predicted V$_{rms}$ to the model. After the best fit was found we determined the range of M$_\bullet$ within $3\sigma$ level ($\Delta \chi^2=9$). Figure~\ref{f:psfplot} illustrates this process. The symbols represent the V$_{rms}$ of the galaxy model which are fit with Jeans models. The model with the best fit M$_\bullet$ is shown with a dashed line, while the two lines above and below the best fit model are the models that are at $\Delta \chi^2=9$ level away from the best fit $M_\bullet$. For each PSF we calculate the difference in values of $M_\bullet$ for these two models.  This number, expressed at a fraction of the input M$_\bullet$, is an estimate of the accuracy to which the M$_\bullet$ can be determined given the PSF properties. 
 
\begin{figure}
        \includegraphics[width=\columnwidth, bb=30 30 450 590]{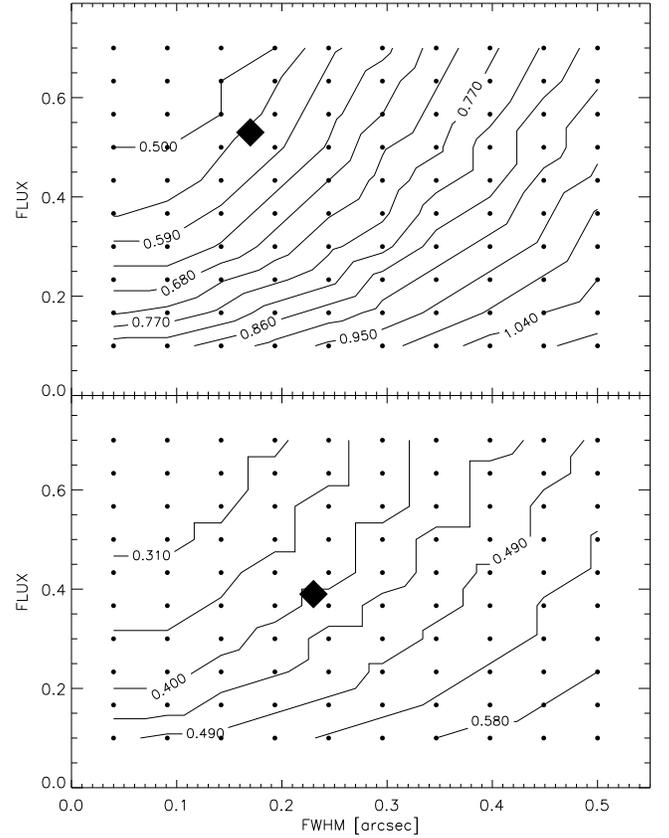}
\caption{\label{f:psfsim} Uncertainty of M$_\bullet$ determination relative to the flux and size (FWHM) of the narrow component of the PSF in the case of  a model galaxy with SMBH whose R$_{sph}\sim0\farcs13$ {\bf (top)}, and a model galaxy with SMBH whose R$_{sph}\sim0\farcs27$ {\bf (bottom)}. Solid points show the parameters of the narrow-Gaussian component of the PSF used in construction of Jeans models. Over-plotted lines are contours of the equal uncertainty to which $M_\bullet$ can be determined using the given PSF. The given numbers are the percentage of the true value of the $M_\bullet$ spaced in 4.5\%.  Filled diamonds mark the PSF values for our observations of NGC2549 (top panel) and NGC524 (bottom panel) for comparison. }
\end{figure}

The results for all PSFs are shown in Fig.~\ref{f:psfsim}. As expected, regardless of the size of the SMBH, the best accuracy is achieved for small FWHM and high flux of the narrow-Gaussian component. The size of the SMBH, however, is important for the absolute accuracy: bigger SMBHs will be determined with higher accuracy.  A typical PSF achievable with the LGS systems, such as in this study, of 0\farcs2 FWHM and 0.4 intensity of the narrow component of the PSF, yields an uncertainty of about 40\% for R$_{sph}$=0\farcs27 and an uncertainty of about 70\% for R$_{sph}$=$0\farcs13$. The uncertainties of  our determinations for NGC524 and NGC2549 are approximately 34\% and 54\%, respectively, which are in a good agreement with the predicted uncertainties of 40\% and 55\% from these simple simulations (diamonds in Fig.~\ref{f:psfsim}). 

Real galaxies are neither pure R$^{1/4}$ laws nor spherically symmetric objects; this exercise is a simplistic in many ways, however, it offers an insight to what improvement one can expect from the current LGS AO observations for SMBH studies. In a very good case of a large SMBH, when the PSF correction is approaching diffraction limit and the highest possible Strehl, it can be expected that the overall uncertainty approaches 25-30\%. On the other hand, it is highly encouraging that even masses of small SMBHs can robustly determined for quite low PSF corrections of e.g. $0\farcs4$ and Strehl of $\sim0.2$. In general, it can be expected that for a decent AO correction, a typical error in M$_\bullet$ in a nearby galaxy will be of the order of a factor of 2. 

This simulations suggest that there is an 'instrumental' limit on individual measurements of 30\% which will translate into the intrinsic scatter of  M$_\bullet- \sigma_e$ relation. Note that due to simple assumptions involved, the simulations can only be used as a guideline for the expected relative improvement in M$_\bullet$ accuracy when using AO corrections, and that the percentages given in these simulations are dependent on the assumed kinematic errors. It, however, does suggest that high signal-to-noise integral field data obtained with LGS AO are capable of robustly measuring M$_\bullet$ in nearby galaxies and, hence, determining the extent (low masses), the shape, as well as increasing limit on the accuracy of the intrinsic scatter of M$_\bullet- \sigma$ relation.

\subsection{Nuclear orbital structure and formation scenarios}
\label{ss:dif}

While both NGC524 ad NGC2549 are classified as fast rotators \citep{2007MNRAS.379..401E}, and seem to have a similar shape but viewed at different inclinations, there are some significant difference between them. NGC524 is more massive and has luminosity weighted stellar populations older than NGC2549.  The light profile of NGC524 shows a shallow core at radii smaller than $0\farcs3$, and of NGC2549 a cusp at the HST resolution. Masses of the black holes are different, and our models suggest the orbital structures around them are also different. These characteristics point out to a different evolution of these objects. 

It has been suggested that the existence of cores in light profiles of massive galaxies is connected to the formation process of nuclei, where high nuclear densities are destroyed in gas poor mergers \citep{1997AJ....114.1771F,2009ApJS..181..486H}. The main point here is that during collisions of galaxies without significant amount of gas, the process of black hole merging removes a significant number of stars with orbits intersecting the newly formed black hole binary.  The binary shrinks ejecting stars mostly on radial orbits that approach the black holes and only stars on tangential orbits remain in a depleted nucleus \citep{2001ApJ...563...34M,2002MNRAS.331L..51M} This process goes in several steps where both classical and relativistic effects have to be taken into account \citep[e.g.][]{2003ApJ...596..860M, 2007ApJ...671...53M}  A different scenario suggests that mergers of galaxies with significant amount of gas lead to a creation of cuspy nuclear light profiles, which are created as a consequence of a nuclear starburst \citep{1994ApJ...437L..47M, 2009ApJS..181..486H}. To some extent this latter scenario includes accretion of cold intergalactic gas. These processes are summarised and discussed in more details in \citet{2008arXiv0810.1681K}  who put forward a hypothesis that "the last major merger that made core ellipticals was gas poor, whereas the last major merger that made coreless ellipticals was gas rich and included a substantial starburst". 

In all these respects the growth of black holes is related to the formation of the nuclei in early-type galaxies. It is somewhat unfortunate that both growth through binary mergers and accretion of gas leave similar imprints in the orbital structure: a tangentially biased anisotropy of the velocity ellipsoid. The difference between the scenarios, however, could be found in the specific angular momentum content of nuclei, if gas poor merges are also responsible for producing slow rotators, while gas rich mergers create fast rotators \citep[e.g.][]{2007MNRAS.379..401E}. If this is true, than slow rotators with cores should have tangential anisotropy in their nuclei. In that case, products of gas poor mergers are slow rotating galaxies, with light deficits and tangential anisotropy in the nuclear orbital distributions (within the core). Given that NGC524 is a fast rotator with these characteristics, it is clear that there is no a simple formation scenario. Finding, however, a slow rotator without a core, or tangentially biased orbits, will argue against gas poor merger scenario as outlined above. Given that kinematically decoupled cores with relatively higher specific angular momenta are often present in slow rotators \citep{2008MNRAS.390...93K}, it would be interesting to see the orbital structure in the vicinity of their central black holes. 

Within present scenarios, the fact that NGC524 has a core and very old stellar population, but also a prominent dusty disk, and has evidence of being a spatially constrained post-starburst galaxy  (Kuntschner et al. in prep, Shapiro et al. in prep), suggest that in this galaxy first a core was created (i.e through a binary black hole merger), but soon after that accretion of gas created the disk and specified the angular momentum of the galaxy, while also possibly contributing to the final growth of the black hole, keeping or enhancing the nuclear tangential anisotropy.

%
%
\section{Conclusions}
\label{s:conc}

In this work we presented observations of two early-type galaxies (NGC524 and NGC2549) with laser guide star adaptive optics obtained at GEMINI North telescope using the NIFS integral field spectrograph. The purpose of these observations was to measure the kinematics within the nuclei of these galaxies and, in combination with previously obtained wide field observations with SAURON at William Herschel Telescope, to determine the masses of central black holes by means of stellar dynamical modeling. A speciality of these observations was that there were no natural guide stars for the selected objects and they were performed using an `open loop' focus model. In this mode, the nucleus of the galaxy is used as a substitute for the natural guide star, but since it is not bright or distinct enough, the variations of the laser guide star focus are not tracked in real time, but varied following a geometric model. We selected two galaxies with different light profiles in order to test the dependence of the AO corrections on the cusp/core nature of the nucleus.

We estimated the seeing correction of the the LGS system by finding the PSF that, when used to convolve an HST/WFPC2 image or deconvolved MGE model, best matched the reconstructed NIFS images. The deconvolved MGE models were useful to verify that the derived PSF estimates were reliable, since the LGS AO observations approach the spatial resolution of HST, making it potentially difficult to neglect the intrinsic HST image quality. We parameterise the PSF with a double-Gaussian, consisting of a narrow and broad component. The final full-width-half-maximum PSFs of our observations are $0\farcs23$ and $1\farcs25$ for NGC524 and $0\farcs17$ and $0\farcs84$ for NGC2549, with the narrow component flux  contribution of 0.39 and 0.53 for NGC524 and NGC2549, respectively. These values demonstrate that the laser correction works well, giving a significant narrow component close to diffraction limit, to some extent regardless of the central light profile. The broad component of the PSF seems to be also dependent on the steepness of the light profile. This can be understood in terms of how much low-order (tip-tilt) correction can be made using the nucleus, which directly depends on the total brightness and the steepness of the nuclear light profile. In addition, the uncertainties in the estimate of the PSF parameters depend on the steepness of the profile.  Our observations confirm that the `open loop' focus model provides a robust and stable method by which a range of objects with faint, extended light profiles can be successfully used as tip-tilt references for LGS AO.  The steeper light profiles, however, will yield a larger contrast within the field-of-view of the wave front sensor, and result in the overall better PSF, making the 'open loop' method more suitable for smaller early-type galaxies.

Our observations were taken in $K$ band, and we use the CO band head to derive the stellar kinematics. We compare these kinematics with the optical and large scale SAURON kinematics and find that they match well. Both data sets are used to constrain three-integral Schwarzschild dynamical models. The models are axisymmetric, numerical and based on an orbit superposition method. For NGC524 they yield M$_{\bullet}$=($8.3^{+2.7}_{-1.5})\times 10^8$ M$_{\sun}$ and for NGC2549 M$_{\bullet}$=($1.4^{+0.2}_{-1.3} )\times 10^7$ M$_{\sun}$ . We showed that large-field IFU data are crucial for constraining the global M/L and contribute to the determination of the M$_{\bullet}$ in cases when black hole's sphere of influence is of similar scales as the obtained spatial resolution.  If only the high-resolution data of small field-of-view are used to constrain the models, the best fitting values are similar, but the uncertainties are considerably larger: up to 2 times for M$_{\bullet}$ and 5 times for M/L. The combination of both large scale (SAURON) and high resolution (NIFS) data is therefore essential to provide useful constraints on M$_{\bullet}$.

The nuclear orbital distribution of the best fitting models is different for these two galaxies. NGC524 has the clear signature of a tangentially anisotropic velocity ellipsoid in the central 0\farcs5, which corresponds to the SMBH sphere of influence, the change in the light profile towards a central core and change in the shape of the LOSVD. In the case of NGC2549 more significant is that  $\sigma_z < \sigma_R$, and  $\sigma_R \approx \sigma_\phi$ over the whole field-of-view covered with the kinematic data. The difference in the nuclear dynamics between these two galaxies suggest that they had different formation processes, likely linked with the growth of the central black holes. The lack of tangential anisotropy next to the SMBH in NGC2549 could, however, be a consequence of barely resolving its sphere of influence. 

We constructed two spherical Jeans models of galaxies with SMBHs of different masses and different spheres of influence and explored a relative improvement one should expect in determinations of M$_\bullet$ in these galaxies given a range in seeing parameters typical for present LGS AO system. We find that for PSF approaching diffraction limit and high Strehl ratios the uncertainty in M$_\bullet$ approaches 25-30\%, but for a more typical PSF it is about 50\%. Our results suggest that systematic studies of SMBHs in the nearby galaxies with LGS AO are able to decrease the intrinsic scatter and determine the extent and shape of M$_\bullet  - \sigma_e$ relation.

\vspace{+1cm}
\noindent{\bf Acknowledgements}\\

\noindent Authors thank Chad Trujillo and the Gemini staff on the help in establishing the 'open loop' observational method and general support of the program.  DK thanks Marc Sarzi for help in the early stages of this project, Kristen Shapiro for providing the code for PSF comparison and Nicholas Scott for the help with MGE model of NGC2549.  DK acknowledges the support from Queen's College Oxford and hospitality of Centre for Astrophysics Research at the University of  Hertfordshire. MC acknowledges support from a STFC Advanced Fellowship (PP/D005574/1). This paper is based on observations obtained at the William Herschel Telescope, operated by the Isaac Newton Group in the Spanish Observatorio del Roque de los Muchachos of the Instituto de Astrof\'{\i}sica de Canarias.  Part of this work is based on data obtained from the ESO/ST-ECF Science Archive Facility. Photometric data were obtained (in part) using the 1.3m McGraw-Hill telescope of the MDM Observatory.NSO/Kitt Peak FTS data used here were produced by NSF/NOAO. Based on observations obtained at the Gemini Observatory, which is operated by the Association of Universities for Research in Astronomy, Inc., under a cooperative agreement with the NSF on behalf of the Gemini partnership: the National Science Foundation (United States), the Science and Technology Facilities Council (United Kingdom), the National Research Council (Canada), CONICYT (Chile), the Australian Research Council (Australia), MinistŽrio da Cincia e Tecnologia (Brazil) and Ministerio de Ciencia, Tecnolog'a e Innovaci—n Productiva  (Argentina). Based on observations made with the NASA/ESA Hubble Space Telescope, obtained [from the Data Archive] at the Space Telescope Science Institute, which is operated by the Association of Universities for Research in Astronomy, Inc., under NASA contract NAS 5-26555. This research has made use of the NASA/IPAC Extragalactic Database (NED) which is operated by the Jet Propulsion Laboratory, California Institute of Technology, under contract with the National Aeronautics and Space Administration.

.


\begin{thebibliography}{}

\bibitem[\protect\citeauthoryear{{Aller} \& {Richstone}}{{Aller} \&
  {Richstone}}{2007}]{2007ApJ...665..120A}
{Aller} M.~C.,  {Richstone} D.~O.,  2007, \apj, 665, 120

\bibitem[\protect\citeauthoryear{{Bacon}, {Copin}, {Monnet}, {Miller},
  {Allington-Smith}, {Bureau}, {Carollo}, {Davies}, {Emsellem}, {Kuntschner},
  {Peletier}, {Verolme} \& {de Zeeuw}}{{Bacon}
  et~al.}{2001}]{2001MNRAS.326...23B}
{Bacon} R.,  {Copin} Y.,  {Monnet} G.,  {Miller} B.~W.,  {Allington-Smith}
  J.~R.,  {Bureau} M.,  {Carollo} C.~M.,  {Davies} R.~L.,  {Emsellem} E.,
  {Kuntschner} H.,  {Peletier} R.~F.,  {Verolme} E.~K.,    {de Zeeuw} P.~T.,
  2001, \mnras, 326, 23

\bibitem[\protect\citeauthoryear{{Bender}, {Saglia} \& {Gerhard}}{{Bender}
  et~al.}{1994}]{1994MNRAS.269..785B}
{Bender} R.,  {Saglia} R.~P.,    {Gerhard} O.~E.,  1994, \mnras, 269, 785

\bibitem[\protect\citeauthoryear{{Binney} \& {Merrifield}}{{Binney} \&
  {Merrifield}}{1998}]{1998gaas.book.....B}
{Binney} J.,  {Merrifield} M.,  1998, {Galactic astronomy}.
Princeton, NJ, Princeton University Press

\bibitem[\protect\citeauthoryear{{Boccas}, {Rigaut}, {Bec}, {Irarrazaval},
  {James}, {Ebbers}, {d'Orgeville}, {Grace}, {Arriagada}, {Karewicz},
  {Sheehan}, {White} \& {Chan}}{{Boccas} et~al.}{2006}]{2006SPIE.6272E.114B}
{Boccas} M.,  {Rigaut} F.,  {Bec} M.,  {Irarrazaval} B.,  {James} E.,  {Ebbers}
  A.,  {d'Orgeville} C.,  {Grace} K.,  {Arriagada} G.,  {Karewicz} S.,
  {Sheehan} M.,  {White} J.,    {Chan} S.,  2006, in Society of Photo-Optical
  Instrumentation Engineers (SPIE) Conference Series Vol.~6272 of Society of
  Photo-Optical Instrumentation Engineers (SPIE) Conference Series, {Laser
  guide star upgrade of Altair at Gemini North}

\bibitem[\protect\citeauthoryear{{Bolton}, {Treu}, {Koopmans}, {Gavazzi},
  {Moustakas}, {Burles}, {Schlegel} \& {Wayth}}{{Bolton}
  et~al.}{2008}]{2008ApJ...684..248B}
{Bolton} A.~S.,  {Treu} T.,  {Koopmans} L.~V.~E.,  {Gavazzi} R.,  {Moustakas}
  L.~A.,  {Burles} S.,  {Schlegel} D.~J.,    {Wayth} R.,  2008, \apj, 684, 248

\bibitem[\protect\citeauthoryear{{Bower}, {Benson}, {Malbon}, {Helly}, {Frenk},
  {Baugh}, {Cole} \& {Lacey}}{{Bower} et~al.}{2006}]{2006MNRAS.370..645B}
{Bower} R.~G.,  {Benson} A.~J.,  {Malbon} R.,  {Helly} J.~C.,  {Frenk} C.~S.,
  {Baugh} C.~M.,  {Cole} S.,    {Lacey} C.~G.,  2006, \mnras, 370, 645

\bibitem[\protect\citeauthoryear{{Cappellari}}{{Cappellari}}{2002}]{2002MNRAS.%
333..400C}
{Cappellari} M.,  2002, \mnras, 333, 400

\bibitem[\protect\citeauthoryear{{Cappellari}}{{Cappellari}}{2008}]{2008MNRAS.%
390...71C}
{Cappellari} M.,  2008, \mnras, 390, 71

\bibitem[\protect\citeauthoryear{{Cappellari}, {Bacon}, {Bureau}, {Damen},
  {Davies}, {de Zeeuw}, {Emsellem}, {Falc{\'o}n-Barroso}, {Krajnovi{\'c}},
  {Kuntschner}, {McDermid}, {Peletier}, {Sarzi}, {van den Bosch} \& {van de
  Ven}}{{Cappellari} et~al.}{2006}]{2006MNRAS.366.1126C}
{Cappellari} M.,  {Bacon} R.,  {Bureau} M.,  {Damen} M.~C.,  {Davies} R.~L.,
  {de Zeeuw} P.~T.,  {Emsellem} E.,  {Falc{\'o}n-Barroso} J.,  {Krajnovi{\'c}}
  D.,  {Kuntschner} H.,  {McDermid} R.~M.,  {Peletier} R.~F.,  {Sarzi} M.,
  {van den Bosch} R.~C.~E.,    {van de Ven} G.,  2006, \mnras, 366, 1126

\bibitem[\protect\citeauthoryear{{Cappellari} \& {Copin}}{{Cappellari} \&
  {Copin}}{2003}]{2003MNRAS.342..345C}
{Cappellari} M.,  {Copin} Y.,  2003, \mnras, 342, 345

\bibitem[\protect\citeauthoryear{{Cappellari} \& {Emsellem}}{{Cappellari} \&
  {Emsellem}}{2004}]{2004PASP..116..138C}
{Cappellari} M.,  {Emsellem} E.,  2004, \pasp, 116, 138

\bibitem[\protect\citeauthoryear{{Cappellari}, {Emsellem}, {Bacon}, {Bureau},
  {Davies}, {de Zeeuw}, {Falc{\'o}n-Barroso}, {Krajnovi{\'c}}, {Kuntschner},
  {McDermid}, {Peletier}, {Sarzi}, {van den Bosch} \& {van de
  Ven}}{{Cappellari} et~al.}{2007}]{2007MNRAS.379..418C}
{Cappellari} M.,  {Emsellem} E.,  {Bacon} R.,  {Bureau} M.,  {Davies} R.~L.,
  {de Zeeuw} P.~T.,  {Falc{\'o}n-Barroso} J.,  {Krajnovi{\'c}} D.,
  {Kuntschner} H.,  {McDermid} R.~M.,  {Peletier} R.~F.,  {Sarzi} M.,  {van den
  Bosch} R.~C.~E.,    {van de Ven} G.,  2007, \mnras, 379, 418

\bibitem[\protect\citeauthoryear{{Cappellari} \& {McDermid}}{{Cappellari} \&
  {McDermid}}{2005}]{2005CQGra..22S.347C}
{Cappellari} M.,  {McDermid} R.~M.,  2005, Classical and Quantum Gravity, 22,
  347

\bibitem[\protect\citeauthoryear{{Cappellari}, {Neumayer}, {Reunanen}, {van der
  Werf}, {de Zeeuw} \& {Rix}}{{Cappellari} et~al.}{2009}]{2009MNRAS.394..660C}
{Cappellari} M.,  {Neumayer} N.,  {Reunanen} J.,  {van der Werf} P.~P.,  {de
  Zeeuw} P.~T.,    {Rix} H.-W.,  2009, \mnras, 394, 660

\bibitem[\protect\citeauthoryear{{Copin}, {Cretton} \& {Emsellem}}{{Copin}
  et~al.}{2004}]{2004A&A...415..889C}
{Copin} Y.,  {Cretton} N.,    {Emsellem} E.,  2004, \aap, 415, 889

\bibitem[\protect\citeauthoryear{{Croton}, {Springel}, {White}, {De Lucia},
  {Frenk}, {Gao}, {Jenkins}, {Kauffmann}, {Navarro} \& {Yoshida}}{{Croton}
  et~al.}{2006}]{2006MNRAS.365...11C}
{Croton} D.~J.,  {Springel} V.,  {White} S.~D.~M.,  {De Lucia} G.,  {Frenk}
  C.~S.,  {Gao} L.,  {Jenkins} A.,  {Kauffmann} G.,  {Navarro} J.~F.,
  {Yoshida} N.,  2006, \mnras, 365, 11

\bibitem[\protect\citeauthoryear{{Davies}}{{Davies}}{2007}]{2007astro.ph..3044%
D}
{Davies} R.,  2007, 9 pages, invited contribution to the 2007 ESO Instrument
  Calibration Workshop, arXiv:astro-ph/0703044

\bibitem[\protect\citeauthoryear{{Davies}, {Rabien}, {Lidman}, {Le Louarn},
  {Kasper}, {F{\"o}rster Schreiber}, {Roccatagliata}, {Ageorges}, {Amico},
  {Dumas} \& {Mannucci}}{{Davies} et~al.}{2008}]{2008Msngr.131....7D}
{Davies} R.,  {Rabien} S.,  {Lidman} C.,  {Le Louarn} M.,  {Kasper} M.,
  {F{\"o}rster Schreiber} N.~M.,  {Roccatagliata} V.,  {Ageorges} N.,  {Amico}
  P.,  {Dumas} C.,    {Mannucci} F.,  2008, The Messenger, 131, 7

\bibitem[\protect\citeauthoryear{{Davies}, {Thomas}, {Genzel}, {S{\'a}nchez},
  {Tacconi}, {Sternberg}, {Eisenhauer}, {Abuter}, {Saglia} \&
  {Bender}}{{Davies} et~al.}{2006}]{2006ApJ...646..754D}
{Davies} R.~I.,  {Thomas} J.,  {Genzel} R.,  {S{\'a}nchez} F.~M.,  {Tacconi}
  L.~J.,  {Sternberg} A.,  {Eisenhauer} F.,  {Abuter} R.,  {Saglia} R.,
  {Bender} R.,  2006, \apj, 646, 754

\bibitem[\protect\citeauthoryear{{de Lorenzi}, {Debattista}, {Gerhard} \&
  {Sambhus}}{{de Lorenzi} et~al.}{2007}]{2007MNRAS.376...71D}
{de Lorenzi} F.,  {Debattista} V.~P.,  {Gerhard} O.,    {Sambhus} N.,  2007,
  \mnras, 376, 71

\bibitem[\protect\citeauthoryear{{de Vaucouleurs}, {de Vaucouleurs}, {Corwin}
  Jr., {Buta}, {Paturel} \& {Fouque}}{{de Vaucouleurs}
  et~al.}{1991}]{1991trcb.book.....D}
{de Vaucouleurs} G.,  {de Vaucouleurs} A.,  {Corwin} Jr. H.~G.,  {Buta} R.~J.,
  {Paturel} G.,    {Fouque} P.,  1991, {Third Reference Catalogue of Bright
  Galaxies}.
Volume 1-3, XII, 2069 pp.~7 figs..~ Springer-Verlag Berlin Heidelberg New York

\bibitem[\protect\citeauthoryear{{Di Matteo}, {Springel} \& {Hernquist}}{{Di
  Matteo} et~al.}{2005}]{2005Natur.433..604D}
{Di Matteo} T.,  {Springel} V.,    {Hernquist} L.,  2005, \nat, 433, 604

\bibitem[\protect\citeauthoryear{{Doeleman}, {Shen}, {Rogers}, {Bower},
  {Wright}, {Zhao}, {Backer}, {Crowley}, {Freund}, {Ho}, {Lo} \&
  {Woody}}{{Doeleman} et~al.}{2001}]{2001AJ....121.2610D}
{Doeleman} S.~S.,  {Shen} Z.-Q.,  {Rogers} A.~E.~E.,  {Bower} G.~C.,  {Wright}
  M.~C.~H.,  {Zhao} J.~H.,  {Backer} D.~C.,  {Crowley} J.~W.,  {Freund} R.~W.,
  {Ho} P.~T.~P.,  {Lo} K.~Y.,    {Woody} D.~P.,  2001, \aj, 121, 2610

\bibitem[\protect\citeauthoryear{{Dolphin}}{{Dolphin}}{2000}]{2000PASP..112.13%
97D}
{Dolphin} A.~E.,  2000, \pasp, 112, 1397

\bibitem[\protect\citeauthoryear{{Emsellem}, {Cappellari}, {Krajnovi{\'c}},
  {van de Ven}, {Bacon}, {Bureau}, {Davies}, {de Zeeuw}, {Falc{\'o}n-Barroso},
  {Kuntschner}, {McDermid}, {Peletier} \& {Sarzi}}{{Emsellem}
  et~al.}{2007}]{2007MNRAS.379..401E}
{Emsellem} E.,  {Cappellari} M.,  {Krajnovi{\'c}} D.,  {van de Ven} G.,
  {Bacon} R.,  {Bureau} M.,  {Davies} R.~L.,  {de Zeeuw} P.~T.,
  {Falc{\'o}n-Barroso} J.,  {Kuntschner} H.,  {McDermid} R.,  {Peletier} R.~F.,
     {Sarzi} M.,  2007, \mnras, 379, 401

\bibitem[\protect\citeauthoryear{{Emsellem}, {Cappellari}, {Peletier},
  {McDermid}, {Geacon}, {Bureau}, {Copin}, {Davies}, {Krajnovi{\' c}},
  {Kuntschner}, {Miller} \& {de Zeeuw}}{{Emsellem}
  et~al.}{2004}]{2004MNRAS.352..721E}
{Emsellem} E.,  {Cappellari} M.,  {Peletier} R.~F.,  {McDermid} R.~M.,
  {Geacon} R.,  {Bureau} M.,  {Copin} Y.,  {Davies} R.~L.,  {Krajnovi{\' c}}
  D.,  {Kuntschner} H.,  {Miller} B.~W.,    {de Zeeuw} P.~T.,  2004, \mnras,
  352, 721

\bibitem[\protect\citeauthoryear{{Emsellem}, {Monnet} \& {Bacon}}{{Emsellem}
  et~al.}{1994}]{1994A&A...285..723E}
{Emsellem} E.,  {Monnet} G.,    {Bacon} R.,  1994, \aap, 285, 723

\bibitem[\protect\citeauthoryear{{Faber}, {Tremaine}, {Ajhar}, {Byun},
  {Dressler}, {Gebhardt}, {Grillmair}, {Kormendy}, {Lauer} \&
  {Richstone}}{{Faber} et~al.}{1997}]{1997AJ....114.1771F}
{Faber} S.~M.,  {Tremaine} S.,  {Ajhar} E.~A.,  {Byun} Y.-I.,  {Dressler} A.,
  {Gebhardt} K.,  {Grillmair} C.,  {Kormendy} J.,  {Lauer} T.~R.,
  {Richstone} D.,  1997, \aj, 114, 1771

\bibitem[\protect\citeauthoryear{{Falcke}, {Markoff} \& {Bower}}{{Falcke}
  et~al.}{2009}]{2009arXiv0901.3723F}
{Falcke} H.,  {Markoff} S.,    {Bower} G.~C.,  2009, ArXiv e-prints

\bibitem[\protect\citeauthoryear{{Ferrarese}}{{Ferrarese}}{2002}]{2002ApJ...57%
8...90F}
{Ferrarese} L.,  2002, \apj, 578, 90

\bibitem[\protect\citeauthoryear{{Ferrarese}}{{Ferrarese}}{2003}]{2003ASPC..29%
1..196F}
{Ferrarese} L.,  2003, in {Sembach} K.~R.,  {Blades} J.~C.,  {Illingworth}
  G.~D.,   {Kennicutt} Jr. R.~C.,  eds, Hubble's Science Legacy: Future
  Optical/Ultraviolet Astronomy from Space Vol.~291 of Astronomical Society of
  the Pacific Conference Series, {Supermassive Black Hole Research in the
  Post-HST Era}.
pp 196--+

\bibitem[\protect\citeauthoryear{{Ferrarese} \& {Ford}}{{Ferrarese} \&
  {Ford}}{2005}]{2005SSRv..116..523F}
{Ferrarese} L.,  {Ford} H.,  2005, Space Science Reviews, 116, 523

\bibitem[\protect\citeauthoryear{{Ferrarese} \& {Merritt}}{{Ferrarese} \&
  {Merritt}}{2000}]{2000ApJ...539L...9F}
{Ferrarese} L.,  {Merritt} D.,  2000, \apjl, 539, L9

\bibitem[\protect\citeauthoryear{{Ferrarese}, {van den Bosch}, {Ford}, {Jaffe}
  \& {O'Connell}}{{Ferrarese} et~al.}{1994}]{1994AJ....108.1598F}
{Ferrarese} L.,  {van den Bosch} F.~C.,  {Ford} H.~C.,  {Jaffe} W.,
  {O'Connell} R.~W.,  1994, \aj, 108, 1598

\bibitem[\protect\citeauthoryear{{Franx}}{{Franx}}{1988}]{1988MNRAS.231..285F}
{Franx} M.,  1988, \mnras, 231, 285

\bibitem[\protect\citeauthoryear{{Freedman}, {Madore}, {Gibson}, {Ferrarese},
  {Kelson}, {Sakai}, {Mould}, {Kennicutt}, {Ford}, {Graham}, {Huchra},
  {Hughes}, {Illingworth}, {Macri} \& {Stetson}}{{Freedman}
  et~al.}{2001}]{2001ApJ...553...47F}
{Freedman} W.~L.,  {Madore} B.~F.,  {Gibson} B.~K.,  {Ferrarese} L.,  {Kelson}
  D.~D.,  {Sakai} S.,  {Mould} J.~R.,  {Kennicutt} R.~C.,  {Ford} H.~C.,
  {Graham} J.~A.,  {Huchra} J.~P.,  {Hughes} S.~M.~G.,  {Illingworth} G.~D.,
  {Macri} L.~M.,    {Stetson} P.~B.,  2001, \apj, 553, 47

\bibitem[\protect\citeauthoryear{{Gebhardt}, {Bender}, {Bower}, {Dressler},
  {Faber}, {Filippenko}, {Green}, {Grillmair}, {Ho}, {Kormendy}, {Lauer},
  {Magorrian}, {Pinkney}, {Richstone} \& {Tremaine}}{{Gebhardt}
  et~al.}{2000}]{2000ApJ...539L..13G}
{Gebhardt} K.,  {Bender} R.,  {Bower} G.,  {Dressler} A.,  {Faber} S.~M.,
  {Filippenko} A.~V.,  {Green} R.,  {Grillmair} C.,  {Ho} L.~C.,  {Kormendy}
  J.,  {Lauer} T.~R.,  {Magorrian} J.,  {Pinkney} J.,  {Richstone} D.,
  {Tremaine} S.,  2000, \apjl, 539, L13

\bibitem[\protect\citeauthoryear{{Gebhardt}, {Richstone}, {Tremaine}, {Lauer},
  {Bender}, {Bower}, {Dressler}, {Faber}, {Filippenko}, {Green}, {Grillmair},
  {Ho}, {Kormendy}, {Magorrian} \& {Pinkney}}{{Gebhardt}
  et~al.}{2003}]{2003ApJ...583...92G}
{Gebhardt} K.,  {Richstone} D.,  {Tremaine} S.,  {Lauer} T.~R.,  {Bender} R.,
  {Bower} G.,  {Dressler} A.,  {Faber} S.~M.,  {Filippenko} A.~V.,  {Green} R.,
   {Grillmair} C.,  {Ho} L.~C.,  {Kormendy} J.,  {Magorrian} J.,    {Pinkney}
  J.,  2003, \apj, 583, 92

\bibitem[\protect\citeauthoryear{{Gerhard}, {Kronawitter}, {Saglia} \&
  {Bender}}{{Gerhard} et~al.}{2001}]{2001AJ....121.1936G}
{Gerhard} O.,  {Kronawitter} A.,  {Saglia} R.~P.,    {Bender} R.,  2001, \aj,
  121, 1936

\bibitem[\protect\citeauthoryear{{Gerhard}}{{Gerhard}}{1993}]{1993MNRAS.265..2%
13G}
{Gerhard} O.~E.,  1993, \mnras, 265, 213

\bibitem[\protect\citeauthoryear{{Ghez}, {Duch{\^e}ne}, {Matthews},
  {Hornstein}, {Tanner}, {Larkin}, {Morris}, {Becklin}, {Salim}, {Kremenek},
  {Thompson}, {Soifer}, {Neugebauer} \& {McLean}}{{Ghez}
  et~al.}{2003}]{2003ApJ...586L.127G}
{Ghez} A.~M.,  {Duch{\^e}ne} G.,  {Matthews} K.,  {Hornstein} S.~D.,  {Tanner}
  A.,  {Larkin} J.,  {Morris} M.,  {Becklin} E.~E.,  {Salim} S.,  {Kremenek}
  T.,  {Thompson} D.,  {Soifer} B.~T.,  {Neugebauer} G.,    {McLean} I.,  2003,
  \apjl, 586, L127

\bibitem[\protect\citeauthoryear{{Ghez}, {Salim}, {Weinberg}, {Lu}, {Do},
  {Dunn}, {Matthews}, {Morris}, {Yelda}, {Becklin}, {Kremenek}, {Milosavljevic}
  \& {Naiman}}{{Ghez} et~al.}{2008}]{2008ApJ...689.1044G}
{Ghez} A.~M.,  {Salim} S.,  {Weinberg} N.~N.,  {Lu} J.~R.,  {Do} T.,  {Dunn}
  J.~K.,  {Matthews} K.,  {Morris} M.~R.,  {Yelda} S.,  {Becklin} E.~E.,
  {Kremenek} T.,  {Milosavljevic} M.,    {Naiman} J.,  2008, \apj, 689, 1044

\bibitem[\protect\citeauthoryear{{Gillessen}, {Eisenhauer}, {Trippe},
  {Alexander}, {Genzel}, {Martins} \& {Ott}}{{Gillessen}
  et~al.}{2009}]{2009ApJ...692.1075G}
{Gillessen} S.,  {Eisenhauer} F.,  {Trippe} S.,  {Alexander} T.,  {Genzel} R.,
  {Martins} F.,    {Ott} T.,  2009, \apj, 692, 1075

\bibitem[\protect\citeauthoryear{{Graham}}{{Graham}}{2008}]{2008PASA...25..167%
G}
{Graham} A.~W.,  2008, Publications of the Astronomical Society of Australia,
  25, 167

\bibitem[\protect\citeauthoryear{{Graham}, {Erwin}, {Caon} \&
  {Trujillo}}{{Graham} et~al.}{2001}]{2001ApJ...563L..11G}
{Graham} A.~W.,  {Erwin} P.,  {Caon} N.,    {Trujillo} I.,  2001, \apjl, 563,
  L11

\bibitem[\protect\citeauthoryear{{G{\"u}ltekin}, {Richstone}, {Gebhardt},
  {Lauer}, {Pinkney}, {Aller}, {Bender}, {Dressler}, {Faber}, {Filippenko},
  {Green}, {Ho}, {Kormendy} \& {Siopis}}{{G{\"u}ltekin}
  et~al.}{2009}]{2009ApJ...695.1577G}
{G{\"u}ltekin} K.,  {Richstone} D.~O.,  {Gebhardt} K.,  {Lauer} T.~R.,
  {Pinkney} J.,  {Aller} M.~C.,  {Bender} R.,  {Dressler} A.,  {Faber} S.~M.,
  {Filippenko} A.~V.,  {Green} R.,  {Ho} L.~C.,  {Kormendy} J.,    {Siopis} C.,
   2009, \apj, 695, 1577

\bibitem[\protect\citeauthoryear{{G{\"u}ltekin}, {Richstone}, {Gebhardt},
  {Lauer}, {Tremaine}, {Aller}, {Bender}, {Dressler}, {Faber}, {Filippenko},
  {Green}, {Ho}, {Kormendy}, {Magorrian}, {Pinkney} \& {Siopis}}{{G{\"u}ltekin}
  et~al.}{2009}]{2009ApJ...698..198G}
{G{\"u}ltekin} K.,  {Richstone} D.~O.,  {Gebhardt} K.,  {Lauer} T.~R.,
  {Tremaine} S.,  {Aller} M.~C.,  {Bender} R.,  {Dressler} A.,  {Faber} S.~M.,
  {Filippenko} A.~V.,  {Green} R.,  {Ho} L.~C.,  {Kormendy} J.,  {Magorrian}
  J.,  {Pinkney} J.,    {Siopis} C.,  2009, \apj, 698, 198

\bibitem[\protect\citeauthoryear{{H{\"a}ring} \& {Rix}}{{H{\"a}ring} \&
  {Rix}}{2004}]{2004ApJ...604L..89H}
{H{\"a}ring} N.,  {Rix} H.-W.,  2004, \apjl, 604, L89

\bibitem[\protect\citeauthoryear{{H{\"a}ring-Neumayer}, {Cappellari}, {Rix},
  {Hartung}, {Prieto}, {Meisenheimer} \& {Lenzen}}{{H{\"a}ring-Neumayer}
  et~al.}{2006}]{2006ApJ...643..226H}
{H{\"a}ring-Neumayer} N.,  {Cappellari} M.,  {Rix} H.-W.,  {Hartung} M.,
  {Prieto} M.~A.,  {Meisenheimer} K.,    {Lenzen} R.,  2006, \apj, 643, 226

\bibitem[\protect\citeauthoryear{{Hopkins}, {Hernquist}, {Cox}, {Di Matteo},
  {Robertson} \& {Springel}}{{Hopkins} et~al.}{2006}]{2006ApJS..163....1H}
{Hopkins} P.~F.,  {Hernquist} L.,  {Cox} T.~J.,  {Di Matteo} T.,  {Robertson}
  B.,    {Springel} V.,  2006, \apjs, 163, 1

\bibitem[\protect\citeauthoryear{{Hopkins}, {Lauer}, {Cox}, {Hernquist} \&
  {Kormendy}}{{Hopkins} et~al.}{2009}]{2009ApJS..181..486H}
{Hopkins} P.~F.,  {Lauer} T.~R.,  {Cox} T.~J.,  {Hernquist} L.,    {Kormendy}
  J.,  2009, \apjs, 181, 486

\bibitem[\protect\citeauthoryear{{Houghton}, {Magorrian}, {Sarzi}, {Thatte},
  {Davies} \& {Krajnovi{\'c}}}{{Houghton} et~al.}{2006}]{2006MNRAS.367....2H}
{Houghton} R.~C.~W.,  {Magorrian} J.,  {Sarzi} M.,  {Thatte} N.,  {Davies}
  R.~L.,    {Krajnovi{\'c}} D.,  2006, \mnras, 367, 2

\bibitem[\protect\citeauthoryear{{Koopmans}, {Treu}, {Bolton}, {Burles} \&
  {Moustakas}}{{Koopmans} et~al.}{2006}]{2006ApJ...649..599K}
{Koopmans} L.~V.~E.,  {Treu} T.,  {Bolton} A.~S.,  {Burles} S.,    {Moustakas}
  L.~A.,  2006, \apj, 649, 599

\bibitem[\protect\citeauthoryear{{Kormendy} \& {Bender}}{{Kormendy} \&
  {Bender}}{2009}]{2009ApJ...691L.142K}
{Kormendy} J.,  {Bender} R.,  2009, \apjl, 691, L142

\bibitem[\protect\citeauthoryear{{Kormendy}, {Fisher}, {Cornell} \&
  {Bender}}{{Kormendy} et~al.}{2008}]{2008arXiv0810.1681K}
{Kormendy} J.,  {Fisher} D.~B.,  {Cornell} M.~E.,    {Bender} R.,  2008, ArXiv
  e-prints

\bibitem[\protect\citeauthoryear{{Kormendy} \& {Richstone}}{{Kormendy} \&
  {Richstone}}{1995}]{1995ARA&A..33..581K}
{Kormendy} J.,  {Richstone} D.,  1995, \araa, 33, 581

\bibitem[\protect\citeauthoryear{{Krajnovi{\' c}}, {Cappellari}, {Emsellem},
  {McDermid} \& {de Zeeuw}}{{Krajnovi{\' c}}
  et~al.}{2005}]{2005MNRAS.357.1113K}
{Krajnovi{\' c}} D.,  {Cappellari} M.,  {Emsellem} E.,  {McDermid} R.~M.,
  {de Zeeuw} P.~T.,  2005, \mnras, 357, 1113

\bibitem[\protect\citeauthoryear{{Krajnovi{\'c}}, {Bacon}, {Cappellari},
  {Davies}, {de Zeeuw}, {Emsellem}, {Falc{\'o}n-Barroso}, {Kuntschner},
  {McDermid}, {Peletier}, {Sarzi}, {van den Bosch} \& {van de
  Ven}}{{Krajnovi{\'c}} et~al.}{2008}]{2008MNRAS.390...93K}
{Krajnovi{\'c}} D.,  {Bacon} R.,  {Cappellari} M.,  {Davies} R.~L.,  {de Zeeuw}
  P.~T.,  {Emsellem} E.,  {Falc{\'o}n-Barroso} J.,  {Kuntschner} H.,
  {McDermid} R.~M.,  {Peletier} R.~F.,  {Sarzi} M.,  {van den Bosch} R.~C.~E.,
    {van de Ven} G.,  2008, \mnras, 390, 93

\bibitem[\protect\citeauthoryear{{Krist} \& {Hook}}{{Krist} \&
  {Hook}}{2001}]{TinyTim}
{Krist} J.,  {Hook} R.,  2001, The Tiny Tim User's Manual, version 6.0

\bibitem[\protect\citeauthoryear{{Kuntschner}, {Emsellem}, {Bacon}, {Bureau},
  {Cappellari}, {Davies}, {de Zeeuw}, {Falc{\'o}n-Barroso}, {Krajnovi{\'c}},
  {McDermid}, {Peletier} \& {Sarzi}}{{Kuntschner}
  et~al.}{2006}]{2006MNRAS.369..497K}
{Kuntschner} H.,  {Emsellem} E.,  {Bacon} R.,  {Bureau} M.,  {Cappellari} M.,
  {Davies} R.~L.,  {de Zeeuw} P.~T.,  {Falc{\'o}n-Barroso} J.,  {Krajnovi{\'c}}
  D.,  {McDermid} R.~M.,  {Peletier} R.~F.,    {Sarzi} M.,  2006, \mnras, 369,
  497

\bibitem[\protect\citeauthoryear{{Kurucz}}{{Kurucz}}{1991}]{1991ppag.proc...27%
K}
{Kurucz} R.~L.,  1991, in {Philip} A.~G.~D.,  {Upgren} A.~R.,   {Janes} K.~A.,
  eds, Precision Photometry: Astrophysics of the Galaxy {New Lines, New Models,
  New Colors}.
pp 27--+

\bibitem[\protect\citeauthoryear{{Lauer}, {Gebhardt}, {Faber}, {Richstone},
  {Tremaine}, {Kormendy}, {Aller}, {Bender}, {Dressler}, {Filippenko}, {Green}
  \& {Ho}}{{Lauer} et~al.}{2007}]{2007ApJ...664..226L}
{Lauer} T.~R.,  {Gebhardt} K.,  {Faber} S.~M.,  {Richstone} D.,  {Tremaine} S.,
   {Kormendy} J.,  {Aller} M.~C.,  {Bender} R.,  {Dressler} A.,  {Filippenko}
  A.~V.,  {Green} R.,    {Ho} L.~C.,  2007, \apj, 664, 226

\bibitem[\protect\citeauthoryear{{Lawson} \& {Hanson}}{{Lawson} \&
  {Hanson}}{1974}]{1974slsp.book.....L}
{Lawson} C.~L.,  {Hanson} R.~J.,  1974, {Solving least squares problems}.
Prentice-Hall Series in Automatic Computation, Englewood Cliffs: Prentice-Hall,
  1974

\bibitem[\protect\citeauthoryear{{Le Louarn}, {Foy}, {Hubin} \& {Tallon}}{{Le
  Louarn} et~al.}{1998}]{1998MNRAS.295..756L}
{Le Louarn} M.,  {Foy} R.,  {Hubin} N.,    {Tallon} M.,  1998, \mnras, 295, 756

\bibitem[\protect\citeauthoryear{{Livingston} \& {Wallace}}{{Livingston} \&
  {Wallace}}{1991}]{1991aass.book.....L}
{Livingston} W.,  {Wallace} L.,  1991, {An atlas of the solar spectrum in the
  infrared from 1850 to 9000 cm-1 (1.1 to 5.4 micrometer)}.
NSO Technical Report, Tucson: National Solar Observatory, National Optical
  Astronomy Observatory, 1991

\bibitem[\protect\citeauthoryear{{Magorrian}, {Tremaine}, {Richstone},
  {Bender}, {Bower}, {Dressler}, {Faber}, {Gebhardt}, {Green}, {Grillmair},
  {Kormendy} \& {Lauer}}{{Magorrian} et~al.}{1998}]{1998AJ....115.2285M}
{Magorrian} J.,  {Tremaine} S.,  {Richstone} D.,  {Bender} R.,  {Bower} G.,
  {Dressler} A.,  {Faber} S.~M.,  {Gebhardt} K.,  {Green} R.,  {Grillmair} C.,
  {Kormendy} J.,    {Lauer} T.,  1998, \aj, 115, 2285

\bibitem[\protect\citeauthoryear{{Maiolino}, {Rieke} \& {Rieke}}{{Maiolino}
  et~al.}{1996}]{1996AJ....111..537M}
{Maiolino} R.,  {Rieke} G.~H.,    {Rieke} M.~J.,  1996, \aj, 111, 537

\bibitem[\protect\citeauthoryear{{Makino} \& {Ebisuzaki}}{{Makino} \&
  {Ebisuzaki}}{1996}]{1996ApJ...465..527M}
{Makino} J.,  {Ebisuzaki} T.,  1996, \apj, 465, 527

\bibitem[\protect\citeauthoryear{{Marconi} \& {Hunt}}{{Marconi} \&
  {Hunt}}{2003}]{2003ApJ...589L..21M}
{Marconi} A.,  {Hunt} L.~K.,  2003, \apjl, 589, L21

\bibitem[\protect\citeauthoryear{{McDermid}, {Emsellem}, {Shapiro}, {Bacon},
  {Bureau}, {Cappellari}, {Davies}, {de Zeeuw}, {Falc{\'o}n-Barroso},
  {Krajnovi{\'c}}, {Kuntschner}, {Peletier} \& {Sarzi}}{{McDermid}
  et~al.}{2006}]{2006MNRAS.373..906M}
{McDermid} R.~M.,  {Emsellem} E.,  {Shapiro} K.~L.,  {Bacon} R.,  {Bureau} M.,
  {Cappellari} M.,  {Davies} R.~L.,  {de Zeeuw} T.,  {Falc{\'o}n-Barroso} J.,
  {Krajnovi{\'c}} D.,  {Kuntschner} H.,  {Peletier} R.~F.,    {Sarzi} M.,
  2006, \mnras, 373, 906

\bibitem[\protect\citeauthoryear{{Merritt}, {Mikkola} \& {Szell}}{{Merritt}
  et~al.}{2007}]{2007ApJ...671...53M}
{Merritt} D.,  {Mikkola} S.,    {Szell} A.,  2007, \apj, 671, 53

\bibitem[\protect\citeauthoryear{{Mihos} \& {Hernquist}}{{Mihos} \&
  {Hernquist}}{1994}]{1994ApJ...437L..47M}
{Mihos} J.~C.,  {Hernquist} L.,  1994, \apjl, 437, L47

\bibitem[\protect\citeauthoryear{{Milosavljevi{\'c}} \&
  {Merritt}}{{Milosavljevi{\'c}} \& {Merritt}}{2001}]{2001ApJ...563...34M}
{Milosavljevi{\'c}} M.,  {Merritt} D.,  2001, \apj, 563, 34

\bibitem[\protect\citeauthoryear{{Milosavljevi{\'c}} \&
  {Merritt}}{{Milosavljevi{\'c}} \& {Merritt}}{2003}]{2003ApJ...596..860M}
{Milosavljevi{\'c}} M.,  {Merritt} D.,  2003, \apj, 596, 860

\bibitem[\protect\citeauthoryear{{Milosavljevi{\'c}}, {Merritt}, {Rest} \& {van
  den Bosch}}{{Milosavljevi{\'c}} et~al.}{2002}]{2002MNRAS.331L..51M}
{Milosavljevi{\'c}} M.,  {Merritt} D.,  {Rest} A.,    {van den Bosch} F.~C.,
  2002, \mnras, 331, L51

\bibitem[\protect\citeauthoryear{{Monnet}, {Bacon} \& {Emsellem}}{{Monnet}
  et~al.}{1992}]{1992A&A...253..366M}
{Monnet} G.,  {Bacon} R.,    {Emsellem} E.,  1992, \aap, 253, 366

\bibitem[\protect\citeauthoryear{{Neumayer}, {Cappellari}, {Reunanen}, {Rix},
  {van der Werf}, {de Zeeuw} \& {Davies}}{{Neumayer}
  et~al.}{2007}]{2007ApJ...671.1329N}
{Neumayer} N.,  {Cappellari} M.,  {Reunanen} J.,  {Rix} H.-W.,  {van der Werf}
  P.~P.,  {de Zeeuw} P.~T.,    {Davies} R.~I.,  2007, \apj, 671, 1329

\bibitem[\protect\citeauthoryear{{Nowak}, {Saglia}, {Thomas}, {Bender},
  {Davies} \& {Gebhardt}}{{Nowak} et~al.}{2008}]{2008MNRAS.391.1629N}
{Nowak} N.,  {Saglia} R.~P.,  {Thomas} J.,  {Bender} R.,  {Davies} R.~I.,
  {Gebhardt} K.,  2008, \mnras, 391, 1629

\bibitem[\protect\citeauthoryear{{Nowak}, {Saglia}, {Thomas}, {Bender},
  {Pannella}, {Gebhardt} \& {Davies}}{{Nowak}
  et~al.}{2007}]{2007MNRAS.379..909N}
{Nowak} N.,  {Saglia} R.~P.,  {Thomas} J.,  {Bender} R.,  {Pannella} M.,
  {Gebhardt} K.,    {Davies} R.~I.,  2007, \mnras, 379, 909

\bibitem[\protect\citeauthoryear{{Onken}, {Valluri}, {Peterson}, {Pogge},
  {Bentz}, {Ferrarese}, {Vestergaard}, {Crenshaw}, {Sergeev}, {McHardy},
  {Merritt}, {Bower}, {Heckman} \& {Wandel}}{{Onken}
  et~al.}{2007}]{2007ApJ...670..105O}
{Onken} C.~A.,  {Valluri} M.,  {Peterson} B.~M.,  {Pogge} R.~W.,  {Bentz}
  M.~C.,  {Ferrarese} L.,  {Vestergaard} M.,  {Crenshaw} D.~M.,  {Sergeev}
  S.~G.,  {McHardy} I.~M.,  {Merritt} D.,  {Bower} G.~A.,  {Heckman} T.~M.,
  {Wandel} A.,  2007, \apj, 670, 105

\bibitem[\protect\citeauthoryear{{Press}, {Teukolsky}, {Vetterling} \&
  {Flannery}}{{Press} et~al.}{1992}]{1992nrfa.book.....P}
{Press} W.~H.,  {Teukolsky} S.~A.,  {Vetterling} W.~T.,    {Flannery} B.~P.,
  1992, {Numerical recipes in FORTRAN. The art of scientific computing}.
Cambridge: University Press, |c1992, 2nd ed.

\bibitem[\protect\citeauthoryear{{Quinlan} \& {Hernquist}}{{Quinlan} \&
  {Hernquist}}{1997}]{1997NewA....2..533Q}
{Quinlan} G.~D.,  {Hernquist} L.,  1997, New Astronomy, 2, 533

\bibitem[\protect\citeauthoryear{{Rest}, {van den Bosch}, {Jaffe}, {Tran},
  {Tsvetanov}, {Ford}, {Davies} \& {Schafer}}{{Rest}
  et~al.}{2001}]{2001AJ....121.2431R}
{Rest} A.,  {van den Bosch} F.~C.,  {Jaffe} W.,  {Tran} H.,  {Tsvetanov} Z.,
  {Ford} H.~C.,  {Davies} J.,    {Schafer} J.,  2001, \aj, 121, 2431

\bibitem[\protect\citeauthoryear{{Richstone} \& {Tremaine}}{{Richstone} \&
  {Tremaine}}{1988}]{1988ApJ...327...82R}
{Richstone} D.~O.,  {Tremaine} S.,  1988, \apj, 327, 82

\bibitem[\protect\citeauthoryear{{Rix}, {de Zeeuw}, {Cretton}, {van der Marel}
  \& {Carollo}}{{Rix} et~al.}{1997}]{1997ApJ...488..702R}
{Rix} H.,  {de Zeeuw} P.~T.,  {Cretton} N.,  {van der Marel} R.~P.,
  {Carollo} C.~M.,  1997, \apj, 488, 702

\bibitem[\protect\citeauthoryear{{Rusin}, {Kochanek} \& {Keeton}}{{Rusin}
  et~al.}{2003}]{2003ApJ...595...29R}
{Rusin} D.,  {Kochanek} C.~S.,    {Keeton} C.~R.,  2003, \apj, 595, 29

\bibitem[\protect\citeauthoryear{{Rybicki}}{{Rybicki}}{1987}]{1987IAUS..127..3%
97R}
{Rybicki} G.~B.,  1987, in IAU Symp. 127: Structure and Dynamics of Elliptical
  Galaxies {Deprojection of Galaxies - how much can BE Learned}.
pp 397--+

\bibitem[\protect\citeauthoryear{{S{\'a}nchez-Bl{\'a}zquez}, {Peletier},
  {Jim{\'e}nez-Vicente}, {Cardiel}, {Cenarro}, {Falc{\'o}n-Barroso}, {Gorgas},
  {Selam} \& {Vazdekis}}{{S{\'a}nchez-Bl{\'a}zquez}
  et~al.}{2006}]{2006MNRAS.371..703S}
{S{\'a}nchez-Bl{\'a}zquez} P.,  {Peletier} R.~F.,  {Jim{\'e}nez-Vicente} J.,
  {Cardiel} N.,  {Cenarro} A.~J.,  {Falc{\'o}n-Barroso} J.,  {Gorgas} J.,
  {Selam} S.,    {Vazdekis} A.,  2006, \mnras, 371, 703

\bibitem[\protect\citeauthoryear{{Schlegel}, {Finkbeiner} \&
  {Davis}}{{Schlegel} et~al.}{1998}]{1998ApJ...500..525S}
{Schlegel} D.~J.,  {Finkbeiner} D.~P.,    {Davis} M.,  1998, \apj, 500, 525

\bibitem[\protect\citeauthoryear{{Sch{\"o}del}, {Ott}, {Genzel}, {Hofmann},
  {Lehnert}, {Eckart}, {Mouawad} \& {Alexander}}{{Sch{\"o}del}
  et~al.}{2002}]{2002Natur.419..694S}
{Sch{\"o}del} R.,  {Ott} T.,  {Genzel} R.,  {Hofmann} R.,  {Lehnert} M.,
  {Eckart} A.,  {Mouawad} N.,    {Alexander} e.~a.,  2002, \nat, 419, 694

\bibitem[\protect\citeauthoryear{{Schwarzschild}}{{Schwarzschild}}{1979}]{1979%
ApJ...232..236S}
{Schwarzschild} M.,  1979, \apj, 232, 236

\bibitem[\protect\citeauthoryear{{Shapiro}, {Cappellari}, {de Zeeuw},
  {McDermid}, {Gebhardt}, {van den Bosch} \& {Statler}}{{Shapiro}
  et~al.}{2006}]{2006MNRAS.370..559S}
{Shapiro} K.~L.,  {Cappellari} M.,  {de Zeeuw} T.,  {McDermid} R.~M.,
  {Gebhardt} K.,  {van den Bosch} R.~C.~E.,    {Statler} T.~S.,  2006, \mnras,
  370, 559

\bibitem[\protect\citeauthoryear{{Shen}, {Lo}, {Liang}, {Ho} \& {Zhao}}{{Shen}
  et~al.}{2005}]{2005Natur.438...62S}
{Shen} Z.-Q.,  {Lo} K.~Y.,  {Liang} M.-C.,  {Ho} P.~T.~P.,    {Zhao} J.-H.,
  2005, \nat, 438, 62

\bibitem[\protect\citeauthoryear{{Silk} \& {Rees}}{{Silk} \&
  {Rees}}{1998}]{1998A&A...331L...1S}
{Silk} J.,  {Rees} M.~J.,  1998, \aap, 331, L1

\bibitem[\protect\citeauthoryear{{Silva}, {Kuntschner} \& {Lyubenova}}{{Silva}
  et~al.}{2008}]{2008ApJ...674..194S}
{Silva} D.~R.,  {Kuntschner} H.,    {Lyubenova} M.,  2008, \apj, 674, 194

\bibitem[\protect\citeauthoryear{{Siopis}, {Gebhardt}, {Lauer}, {Kormendy},
  {Pinkney}, {Richstone}, {Faber}, {Tremaine}, {Aller}, {Bender}, {Bower},
  {Dressler}, {Filippenko}, {Green}, {Ho} \& {Magorrian}}{{Siopis}
  et~al.}{2009}]{2009ApJ...693..946S}
{Siopis} C.,  {Gebhardt} K.,  {Lauer} T.~R.,  {Kormendy} J.,  {Pinkney} J.,
  {Richstone} D.,  {Faber} S.~M.,  {Tremaine} S.,  {Aller} M.~C.,  {Bender} R.,
   {Bower} G.,  {Dressler} A.,  {Filippenko} A.~V.,  {Green} R.,  {Ho} L.~C.,
   {Magorrian} J.,  2009, \apj, 693, 946

\bibitem[\protect\citeauthoryear{{Statler}}{{Statler}}{1995}]{1995AJ....109.13%
71S}
{Statler} T.,  1995, \aj, 109, 1371

\bibitem[\protect\citeauthoryear{{Stuik}, {Le Louarn} \& {Quirrenbach}}{{Stuik}
  et~al.}{2004}]{2004SPIE.5490..331S}
{Stuik} R.,  {Le Louarn} M.,    {Quirrenbach} A.,  2004, in {Bonaccini Calia}
  D.,  {Ellerbroek} B.~L.,   {Ragazzoni} R.,  eds, Society of Photo-Optical
  Instrumentation Engineers (SPIE) Conference Series Vol.~5490 of Society of
  Photo-Optical Instrumentation Engineers (SPIE) Conference Series,
  {Generalized sky coverage for adaptive optics and interferometry}.
pp 331--337

\bibitem[\protect\citeauthoryear{{Thomas}, {Saglia}, {Bender}, {Thomas},
  {Gebhardt}, {Magorrian}, {Corsini} \& {Wegner}}{{Thomas}
  et~al.}{2007}]{2007MNRAS.382..657T}
{Thomas} J.,  {Saglia} R.~P.,  {Bender} R.,  {Thomas} D.,  {Gebhardt} K.,
  {Magorrian} J.,  {Corsini} E.~M.,    {Wegner} G.,  2007, \mnras, 382, 657

\bibitem[\protect\citeauthoryear{{Tonry}, {Dressler}, {Blakeslee}, {Ajhar},
  {Fletcher}, {Luppino}, {Metzger} \& {Moore}}{{Tonry}
  et~al.}{2001}]{2001ApJ...546..681T}
{Tonry} J.~L.,  {Dressler} A.,  {Blakeslee} J.~P.,  {Ajhar} E.~A.,  {Fletcher}
  A.~B.,  {Luppino} G.~A.,  {Metzger} M.~R.,    {Moore} C.~B.,  2001, \apj,
  546, 681

\bibitem[\protect\citeauthoryear{{Vacca}, {Cushing} \& {Rayner}}{{Vacca}
  et~al.}{2003}]{2003PASP..115..389V}
{Vacca} W.~D.,  {Cushing} M.~C.,    {Rayner} J.~T.,  2003, \pasp, 115, 389

\bibitem[\protect\citeauthoryear{{Valluri}, {Ferrarese}, {Merritt} \&
  {Joseph}}{{Valluri} et~al.}{2005}]{2005ApJ...628..137V}
{Valluri} M.,  {Ferrarese} L.,  {Merritt} D.,    {Joseph} C.~L.,  2005, \apj,
  628, 137

\bibitem[\protect\citeauthoryear{{van de Ven}, {de Zeeuw} \& {van den
  Bosch}}{{van de Ven} et~al.}{2008}]{2008MNRAS.385..614V}
{van de Ven} G.,  {de Zeeuw} P.~T.,    {van den Bosch} R.~C.~E.,  2008, \mnras,
  385, 614

\bibitem[\protect\citeauthoryear{{van den Bosch} \& {van de Ven}}{{van den
  Bosch} \& {van de Ven}}{2008}]{2008arXiv0811.3474V}
{van den Bosch} R.~C.~E.,  {van de Ven} G.,  2008, ArXiv e-prints

\bibitem[\protect\citeauthoryear{{van den Bosch}, {van de Ven}, {Verolme},
  {Cappellari} \& {de Zeeuw}}{{van den Bosch}
  et~al.}{2008}]{2008MNRAS.385..647V}
{van den Bosch} R.~C.~E.,  {van de Ven} G.,  {Verolme} E.~K.,  {Cappellari} M.,
     {de Zeeuw} P.~T.,  2008, \mnras, 385, 647

\bibitem[\protect\citeauthoryear{{van der Marel}, {Cretton}, {de Zeeuw} \&
  {Rix}}{{van der Marel} et~al.}{1998}]{1998ApJ...493..613V}
{van der Marel} R.~P.,  {Cretton} N.,  {de Zeeuw} P.~T.,    {Rix} H.,  1998,
  \apj, 493, 613

\bibitem[\protect\citeauthoryear{{van der Marel} \& {Franx}}{{van der Marel} \&
  {Franx}}{1993}]{1993ApJ...407..525V}
{van der Marel} R.~P.,  {Franx} M.,  1993, \apj, 407, 525

\bibitem[\protect\citeauthoryear{{Verolme}, {Cappellari}, {Copin}, {van der
  Marel}, {Bacon}, {Bureau}, {Davies}, {Miller} \& {de Zeeuw}}{{Verolme}
  et~al.}{2002}]{2002MNRAS.335..517V}
{Verolme} E.~K.,  {Cappellari} M.,  {Copin} Y.,  {van der Marel} R.~P.,
  {Bacon} R.,  {Bureau} M.,  {Davies} R.~L.,  {Miller} B.~M.,    {de Zeeuw}
  P.~T.,  2002, \mnras, 335, 517

\bibitem[\protect\citeauthoryear{{Wallace} \& {Hinkle}}{{Wallace} \&
  {Hinkle}}{1997}]{1997ApJS..111..445W}
{Wallace} L.,  {Hinkle} K.,  1997, \apjs, 111, 445

\bibitem[\protect\citeauthoryear{{Winge}, {Riffel} \&
  {Storchi-Bergmann}}{{Winge} et~al.}{2008}]{2008RMxAC..32..177W}
{Winge} C.,  {Riffel} R.~A.,    {Storchi-Bergmann} T.,  2008, in Revista
  Mexicana de Astronomia y Astrofisica Conference Series Vol.~32 of Revista
  Mexicana de Astronomia y Astrofisica Conference Series, {The Gemini library
  of late spectral templates for stellar kinematics analysis in the CO 2.3mmu
  region}.
pp 177--177

\end{thebibliography}


\appendix

\section{MGE parameters for NGC2549}
\label{s:mgepara}

\begin{table}
   \caption{Parameters of MGE models for NGC2549}
   \label{t:mge}
$$
  \begin{array}{cccc}
    \hline
    \noalign{\smallskip}

  \multicolumn{4}{c}{$NGC2549$}\\
   $j$ &  $log I$_{j} (L_{\odot}{pc}^{-2}) &log \sigma_{j} $(arcsec)$ &$q$_{j} \\
    &  $L$_{\odot}{pc}^{-2}  &  $arcsec$ &\\
    \noalign{\smallskip} \hline \hline \noalign{\smallskip}
   1 & 5.957     & -1.729   &  0.635\\
   2 & 4.926     & -1.118   &  0.687\\
   3 & 4.291     & -0.750   &  0.584\\
  4  & 4.109     & -0.525   &  0.800\\
  5  & 4.089     & -0.200   &  0.800\\
  6  & 3.937     &  0.121   &  0.720\\
  7  & 3.357     &  0.556   &  0.501\\
  8  & 3.081     &  0.663   &  0.766\\
  9  & 3.286     &  0.774   &  0.256\\
 10 & 2.449     &  1.048   &  0.741\\
 11 & 2.651     &  1.360   &  0.305\\
 12 & 1.961     &  1.717   &  0.298\\
 13 &  0.259   &  1.915   &  0.452\\
 14 & 0.592    &  1.915   &  0.800\\

          \noalign{\smallskip}
    \hline
  \end{array}
$$ 
\end{table}

\label{lastpage}

\end{document}